\documentclass[useAMS,usenatbib]{mn2e}
\usepackage{graphicx,epsfig,psfig}
\def\spose#1{\hbox to 0pt{#1\hss}}
\def\simlt{\mathrel{\spose{\lower 3pt\hbox{$\mathchar"218$}}
     \raise 2.0pt\hbox{$\mathchar"13C$}}}
\def\simgt{\mathrel{\spose{\lower 3pt\hbox{$\mathchar"218$}}
     \raise 2.0pt\hbox{$\mathchar"13E$}}}

\title[The Origin of the Light Distribution in Spiral Galaxies]{The
Origin of the Light Distribution in Spiral Galaxies}
\author[S\'anchez-Bl\'azquez et al.]{P. S\'anchez-Bl\'azquez$^{1,2}$, 
S. Courty$^{1}$, B.K. Gibson$^{1}$, and C.B. Brook$^{1}$\\
$^{1}$Jeremiah Horrocks Institute for Astrophysics and Supercomputing, 
University of Central Lancashire, Preston, PR1~2HE, UK\\
$^{2}$Instituto de Astrof\'{\i}sica de Canarias, 
c/V\'{\i}a L\'actea s/n, E38205, La Laguna, Tenerife, Spain}

\begin{document}

\date{Accepted}

\pagerange{\pageref{firstpage}--\pageref{lastpage}} \pubyear{2009}

\maketitle

\label{firstpage}

\begin{abstract}
We analyse a high-resolution, fully cosmological, hydrodynamical
disc galaxy simulation, to study the source of the
double-exponential light profiles seen in many stellar discs, and
the effects of stellar radial migration upon the spatio-temporal
evolution of both the disc age and metallicity distributions.  We
find a ``break'' in the pure exponential stellar surface brightness
profile, and trace its origin to a sharp decrease in the star
formation per unit surface area, itself produced by a decrease in the gas
volume density due to a warping of the gas disc. Star formation in
the disc continues well beyond the break.  We find that the break is
more pronounced in bluer wavebands. By contrast, we find little or no
break in the mass density profile.  This is, in part, due to the net
radial migration of stars towards the external parts of the
disc. Beyond the break radius, we find that $\sim$60\% of the
resident stars migrated from the inner disc, while $\sim$25\% formed
in situ. Our simulated galaxy also has a minimum in the age profile
at the break radius but, in disagreement with some previous studies,
migration is not the main mechanism producing this shape. In our
simulation, the disc metallicity gradient flattens with time,
consistent with an ``inside-out'' formation scenario. We do not find
any difference in the intensity or the position of the break with
inclination, suggesting that perhaps the differences found in
empirical studies are driven by dust extinction.
\end{abstract}
\begin{keywords}
galaxies: spiral-- galaxies: stellar content -- 
galaxies: kinematic and dynamic -- galaxies: structure -- galaxies: abundances
methods: numerical
\end{keywords}
\section{Introduction}
Since 1940, it has been known that the stellar disc light in spiral 
galaxies decreases exponentially with radius \citep[][]{Patt40, Fre70}.  
This is thought to be the result of the initial angular momentum 
distribution of the gas cloud that collapsed to form the disc 
\citep[][]{FE80, Dal97, Mo98, vdB01, FC01}.  However, 
 since the pioneering work of van~der~Kruit (1979), it has
become clear
that this exponential profile does not extend to arbitrarily large radii 
in the majority of disc galaxies. 
\citet{PT06} and Erwin et al. (2008) showed --
for later and earlier types, respectively -- that, in 
fact, the galaxies for which this happens represent a minority, 
identifying three classes of surface brightness behaviour in the 
outskirts of disc galaxies:  $\sim$60\% of the galaxies show a deficit of light 
with respect to an extrapolated singular exponential (Type II); $\sim$30\% 
show an excess of light in the outskirts respect to a singular 
exponential \citep[][Type III]{EBP05, PT06}; while only $\sim$10\% show pure 
exponential profiles \citep[][Type I]{BH05, PT06, Erw08} out to $\simgt$10 
scale-lengths\footnote{These percentages are very dependent upon the 
morphological class of the galaxy (see Beckman et~al. 2006).}. 
Further departure from the notion of the ``disc-as-exponential'' paradigm
leads to even more elaborate classifications based upon, for example, the 
position of the ``break'' with respect to the outer Lindblad Resonance 
of the bar (for Type~II) or the exact shape  of the surface brightness 
profile in 
the outskirts (for Type~III) \citep[eg.,][]{PT06,Erw08}.

The break radius ($R_{\rm br}$), or the radius at which the light 
profile starts to deviate from a pure exponential is empirically seen to 
occur at $\sim$1.5$-$4.5 disc scale-lengths \citep[][]{PT06}.  Outside 
the break, the surface brightness does not immediately drop to zero, but 
follows a second (different) exponential \citep[][]{dG01, Poh02}.  The 
fact that such breaks are very common, at least in late-type galaxies 
(Beckman et~al. 2006), indicates that they either form easily or survive 
for significant periods of time. The exact causes though, 30 years on 
from the identification of this ``break'' phenomenon, remain unclear. 
The leading candidates can be divided broadly into those related to 
angular momentum conservation and those related to a star formation 
threshold.

In the case of the former, it can be shown that {\it if} angular 
momentum redistribution does not occur in the disc, the outermost 
stellar radius will reflect the maximum value of the angular momentum of 
the baryonic material (that of the protogalaxy) \citep{vdK87}.  However, 
numerical simulations have shown that non-axisymmetric instabilities 
{\it do} drive a substantial redistribution of angular momentum in the 
disc \citep{Deb06} and that, in fact, this can be a responsible agent 
for a break. This is because the angular momentum redistribution leads 
to an increase in both the central densities and disc scale-lengths. As 
the angular momentum redistribution cannot act efficiently to 
arbitrarily large radius, a break in the surface mass density 
distribution results.

Alternatively, breaks may be due to the effect of star formation 
thresholds (Kennicutt 1989).  Within this latter scenario, the break 
radius would be located where the density of the gas is lower than the 
critical value for star formation. The existence of extended UV discs 
\citep[][]{Thil07, GdP07}, the lack of correlation between H$\alpha$ 
``cutoffs'' and optical disc breaks \citep[][]{Poh04,EH06}, and the 
exponential decay of the light outside the break, however, have all been 
used as arguments against this picture. That said, \citet{EH06} showed 
that a double exponential profile {\it may} result from a multicomponent 
star formation prescription, where turbulent compression in the outer 
parts of the disc can allow for cloud formation and star formation 
despite sub-critical densities. More recently, models have shown that 
redistribution of stellar material in the disc can also lead to 
exponential profiles outside the star forming disc's break radius 
\citep[][]{Rok08a}.

Not surprisingly, hybrid models have been suggested which combine 
aspects of both pictures. For example, \citet{vdB01} investigated a 
combination of the collapse and the threshold model, claiming that gas 
breaks are a direct consequence of the angular momentum of the 
protogalaxy while the stellar breaks are determined by a star formation 
threshold.

While each proposed model appears capable of explaining the presence
of disc breaks, it is equally important that any successful hypothesis
also takes into account the fact that: (i) breaks are not seen in {\it
all} galaxies \citep{BT97,Wei01,BH05}; (ii) breaks are seen in discs
of {\it all} morphological types, from S0 \citep{Erw08} to Sm (eg.,
Kregel et~al. 2002\nocite{Kreg02}; PT06) and even in irregular galaxies \citep{HE06}
-- albeit, not with the same
frequency; and (iii) breaks are observed out to redshift z$\sim$1
\citep[][]{Per04, TP, Azz08}. Recent observational studies have now
imposed additional constraints on the models for break formation.  For
example, de~Jong et~al. (2007) have shown that the break position is
independent of stellar age, a result compatible with that of  Bakos
et~al. (2008), who found the position of the break to be independent
of the photometric band. It has also been found that the prominence or intensity
of the break decreases with increasing distance from the galactic
mid-plane \citep{dJ07} and for older stellar populations 
\citep{Bak08, Azz08l}.

Furthermore, the recent work of \citet{Rok08a} has demonstrated the
importance of stellar migrations in defining the final mass density
and surface-brightness density profiles of disc galaxies. The
re-organization of stars in the disc can have important consequences
for the observational constrains used for the chemical evolution
models, as the age-metallicity relation or the metallicity
distribution, as previously shown by several studies
\citep[e.g.][]{Hay06, Hay08, SB08, Rok08b}.

Previous work in this area has concentrated upon somewhat idealised scenarios
of isolated galaxies. 
The advantage of this approach is that higher resolution
can be achieved more readily, with a greater associated exploration
of parameter space.
However, such an approach necessarily carries with it the
disadvantage of not being able to take
into account various processes such as the cosmological infall of
gas and/or the interactions with other galaxies that are intrinsic to the
hierarchical assembly paradigm.  Indeed, 
interactions with
satellites can have an important influence in shaping the disc,
as suggested by simulations \citep[e.g.,][]{Youn07, Pen06} and should
not be ignored.  Our work aims to fill this hole by addressing the
possible physical origins of the two-component surface brightness
profiles, but now within a cosmological context.  We will examine
the likelihood that such profiles are the 
consequence of dynamical processes, star
formation processes, or a combination of the two.

\section[]{Simulation Details}

Our simulated disc was realised using
multi-resolved, cosmological simulations generated with
the N-body, hydrodynamical code {\tt RAMSES} \citep{Tey02} which is based 
upon
an Adaptive Mesh Refinement (AMR) technique. {\tt RAMSES} includes
gravitation, hydrodynamics, radiative cooling, and heating processes,
immersed within a
uniform, Haardt \& Madau (1996) ionising radiation
field. In addition to solving the Euler equations with a net
cooling term, {\tt RAMSES} includes a phenomenological treatment of
star formation, which contributes directly to the chemical enrichment of the
interstellar and intergalactic medium via subsequent Type~II 
supernovae.\footnote{The current implementation of chemistry within
{\tt RAMSES} is restricted to the global metal content ($Z$), under
the assumption of the instantaneous recycling approximation; an 
extension to {\tt RAMSES} which relaxes this approximation and 
includes the effects of Type~Ia supernovae and asymptotic giant
branch stars is currently under development (Few et~al. 2010, in
preparation).}
The spatio-temporal evolution of the gas metal abundance, as well as
the redshift-dependent photoionising background, are consistently
accounted for in the computation of the net cooling rates of the gas
(Courty \& Teyssier 2009, in preparation). This combined contribution
has been fit using the {\tt CLOUDY} photoionisation code 
\citep{Ferland98}.

The star formation prescription within {\tt RAMSES} 
is described in detail by \citet{Dubois08}; only a brief summary is 
provided in what follows.
Star formation is permitted in
cells whose density is higher than a given threshold, according to the
following rate: $\dot{\rho_\ast}=-\rho/t_\ast$, where the star formation
timescale is proportional to the local free-fall time,
$t_\ast=t_0(\rho/\rho_0)^{1/2}$. As in Dubois \& Teyssier's work, we
set this timescale to $t_0=8$~Gyr, with an associated
density threshold for star formation of $\rho_0=0.1$ H cm$^{-3}$. 
Kinetic feedback of supernovae energy to the surrounding
interstellar  medium is
modeled with a blast-wave solution that essentially mimics the
expansion of superbubbles whose radius is fixed to two cells.
Thermal feedback is accounted
for by using a polytropic equation of state in the high-density
regions. The amount of gas then turned into stellar material in each
eligible gas cell is $m_\ast(1+\eta_{SN}+\eta_W)$, where the mass locked
in long-lived stars is $m_\ast$, and $m_\ast(\eta_{SN}+\eta_W)$ is the mass
carried away by the blast-wave, with the wind driving parameter 
$\eta_W=1$, and the mass fraction
of stars recycled into supernovae ejecta $\eta_{SN}=0.1$.
Photometric properties of the resulting stellar populations
were derived by assuming that each star particle was a single
stellar population (SSP), and then applying the models of
Bruzual \& Charlot (2003) with a Salpeter (1955) initial mass
function (IMF) and the Bertelli et~al. (1994) isochrones.

We run our simulation  to $z=0$ in 
a $\Lambda$CDM Universe with the following cosmological parameters, 
$\Omega_{\Lambda}$=0.7, $\Omega_m=0.3$, $\Omega_{\rm b}$=0.045, $\sigma_8$=0.9
and H$_{0}=$73~kms$^{-1}$s$^{-1}$Mpc$^{-1}$.
To carry out cosmological simulations of disc galaxies at the sub-kpc
scale, a multi-resolved approach was adopted. Candidate dark matter
halos were chosen in a low-resolution, large-scale structure
simulation. The size of this $\Lambda$CDM simulation was 
20$h^{-1}$~Mpc, 
contains $128^3$ dark matter particles, and was generated with
{\tt RAMSES} as part of the Horizon 
Project\footnote{\tt http://www.projet-horizon.fr}. Our multi-resolved
simulation used three nested boxes whose initial conditions were centred
on our candidate halo and whose resolution ranged from $128^3$ in the
external part of the computational volume to $512^3$ for the most
nested box. The latter resolution is also the one of the coarse grid
in the central area and the adaptive mesh nature of {\tt RAMSES} was
employed to refine a further seven levels, thereby reaching a 
spatial resolution of 435~pc in the central area by $z$=0.  Dark matter 
particles
have a mass of 6$\times$10$^6$~M$_\odot$ and 
the initial gas mass per cell was $\sim$10$^6$~M$_{\odot}$. 

As noted in Gibson et~al. (2008), the selection of the candidate halo
was made without any preconceptions regarding spin $\lambda$
or triaxiality $T$.  Care was taken though to ensure that the
final halo was not contaminated by 
low-resolution dark matter particles up to at least five virial radii. 
The disc galaxy discussed here sits within a low-spin ($\lambda$=0.02), 
mildly oblate triaxial ($T$=0.32) halo of dynamical
mass 7.6$\times$10$^{11}$~M$_{\odot}$.  For redshifts $z$$<$2.3, the 
rotational axis of the disc could be aligned readily using the 
angular momentum vector of the gas cells; the analysis which 
follows is therefore restricted to $z$$<$2.3.  Further technical details
of the simulation and alignment process are forthcoming
(Courty et~al. 2009, in preparation).

\section{Characteristics of the Disc}

Hydrodynamical simulations of galaxies within the canonical
hierarchical structure assembly framework 
are known to have a number of 
difficulties in reproducing disc-dominated galaxies 
(e.g., Springel \& Hernquist 2003\nocite{SH03}; Abadi et~al. 
2003\nocite{Abad03a};
Bailin et~al. 2005; Okamoto et~al\ 2005\nocite{Okam05}; 
Governato et~al. 2007\nocite{Gov07}; Scannapieco et~al. 2008a\nocite{Scan08a}). A major problem is that baryons
condense early and then transfer a significant fraction of their
angular momentum to the dark matter as the final galaxy assembles
\citep{NW94}.  As a result, these galaxies contain a significant
fraction of their final baryonic mass in a spheroidal-like component
supported primarily by velocity anisotropy, with consequent 
bulge-to-disc ratios in excess of those encountered in late-type spirals
today.
 
\begin{figure}
\resizebox{0.4\textwidth}{!}{\includegraphics[angle=-90]{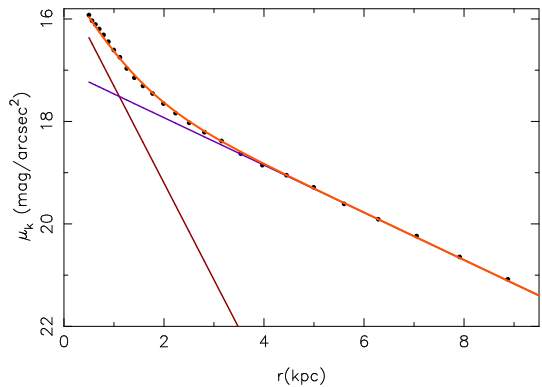}}
\caption{K-band surface brightness profile of our simulated disc, plotted out
to $\sim$3 disc scalelengths.
The fitted profiles of the disc (exponential law, shown in 
purple) and the bulge (Sersic law, shown in dark red) are overlaid, along
with the sum of the two (shown in orange).  \label{decomposition}}
\end{figure}

\begin{figure}
\resizebox{0.4\textwidth}{!}{\includegraphics[angle=-90]{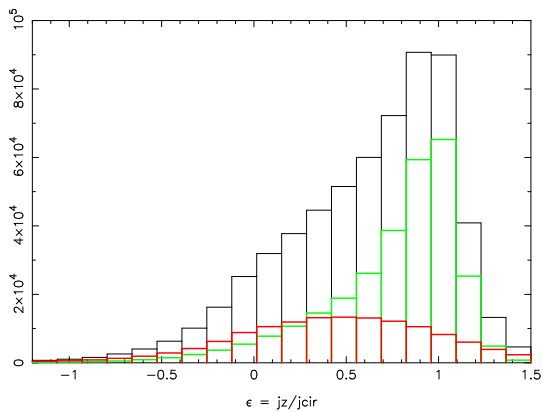}}
\caption{Distribution of $\epsilon=j_z/j_{\rm cir}$ for all stars within a 
sphere of r$<20$ kpc (black line), 
within a sphere of 1.5~kpc (red line), and all those with 3$<$r$<$15 kpc and 
distance from the mid-plane $<$$|$3$|$~kpc (green line).
\label{am.distribution}}
\end{figure}

To derive the bulge-to-disc ratio of our simulated galaxy,
we categorise the stars as ``disc'' or ``spheroid'' following 
\citet{Abad03b} and \citet{Scan08b}.
We derive the distribution of the circularity parameter $\epsilon$ 
(where $\epsilon=j_z/j_{\rm cir}$, and $j_z$ is the $z$-component of
the specific angular momentum
of each star, where the $z$-axis is the symmetry axis and
and j$_{\rm cir}$ is the angular
momentum expected for a circular orbit at the same radius, 
$r$ -- i.e., $j_{\rm cir}=r v_{\rm cir}(r)$).
Figure~\ref{am.distribution} shows the $\epsilon$ distribution of 
all stellar particles within a sphere of radius of 20~kpc at $z$=0 (black).
We have marked, with different colours, those 
particles inside a sphere of radius 1.5~kpc (red), and 
those with radius 3$<r<$15~kpc and within 3~kpc of the disc mid-plane (green).
Considering a stellar disc composed of stars 
with $r$$<$15~kpc, $|z|$$<$3~kpc, and 0.8$<$$\epsilon$$<$1.2, 
and including only those stars rotating in the plane of the disc 
(with $\cos \alpha > $ 0.7, where $\alpha$ is 
the angle between the angular momentum vector of the particle and 
the $z$-axis, after Scannappieco et al.\ 2008b)\nocite{Scan08b},
the disc-to-total ratio in our galaxy -- measured as the ratio between the 
stellar mass on the disc and the total stellar mass -- is 0.37.
Put another way, the bulge-to-disc ratio is, not surprisingly (and in
keeping with the aforementioned particle-based disc simulations), large
with respect to that observed in 
late-type spirals (Sb-Sd), being more consistent with the ratio
expected for an early-type (Sa-S0) galaxy.
Figure~\ref{decomposition} shows a bulge-to-disc decomposition
of our simulation, using the K-band light profile and employing 
an exponential law for the disc
and a Sersic law for the bulge. The fit was done in a iterative 
manner, following MacArthur et al. (2003), resulting in a photometric 
bulge-to-disc ratio of log(B/D)=$-0.5$, consistent with that 
expected for an early-type 
(Sa-S0) galaxy (e.g. morphological type T$\sim$3: 
Simien \& de Vaucouleurs 1986\nocite{SdV86}).

The stellar bulge associated with our simulated disc is partially
supported by rotation (with V$_{\rm rot}$/$\sigma$$\approx$0.5, 
where V$_{\rm rot}$ is the
maximum rotational velocity). In the simulation here, much as for
the Milky Way bulge (admittedly, perhaps just coincidentally), 
V$_{\rm rot}$=70~km~s$^{-1}$. 

\begin{figure*}
\centering
\resizebox{0.3\textwidth}{!}{\includegraphics[angle=-90]{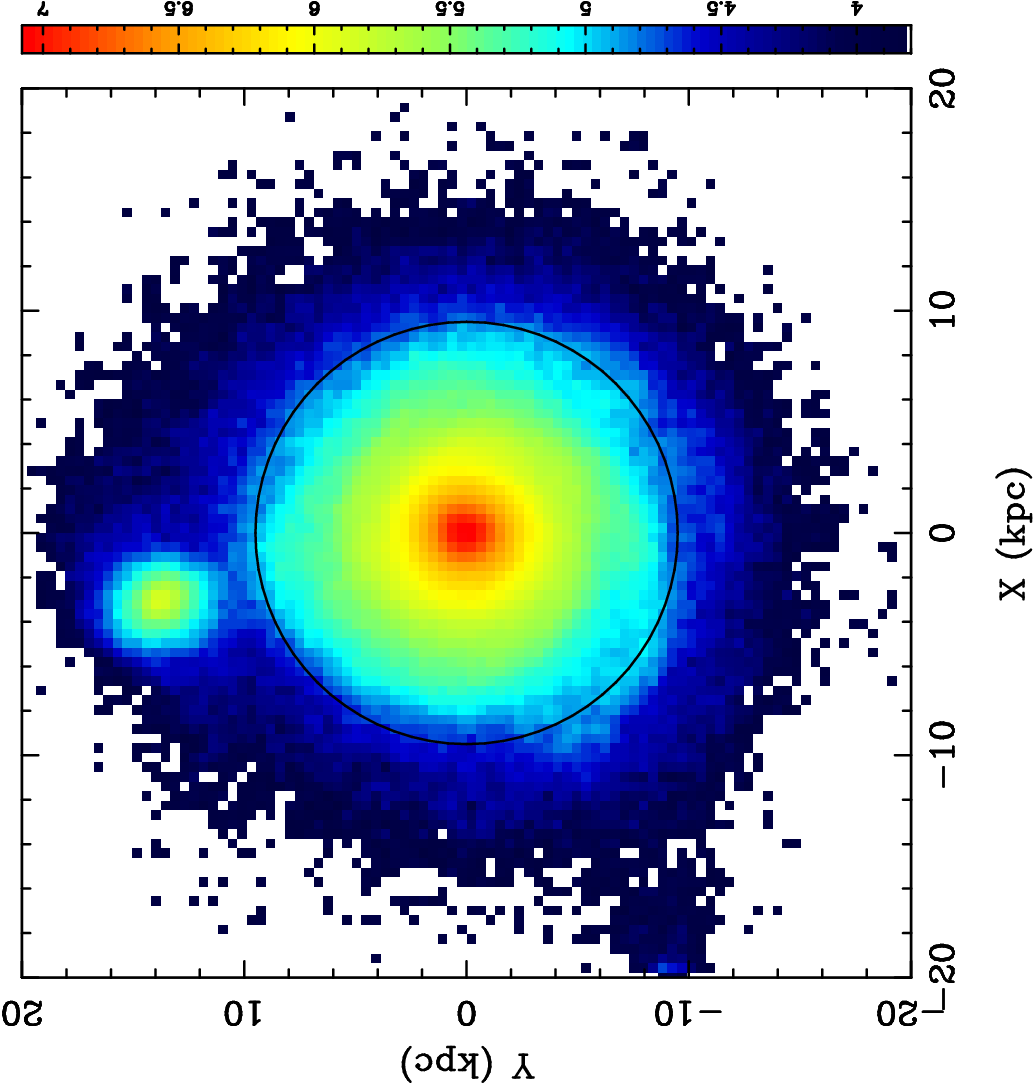}}
\resizebox{0.3\textwidth}{!}{\includegraphics[angle=-90]{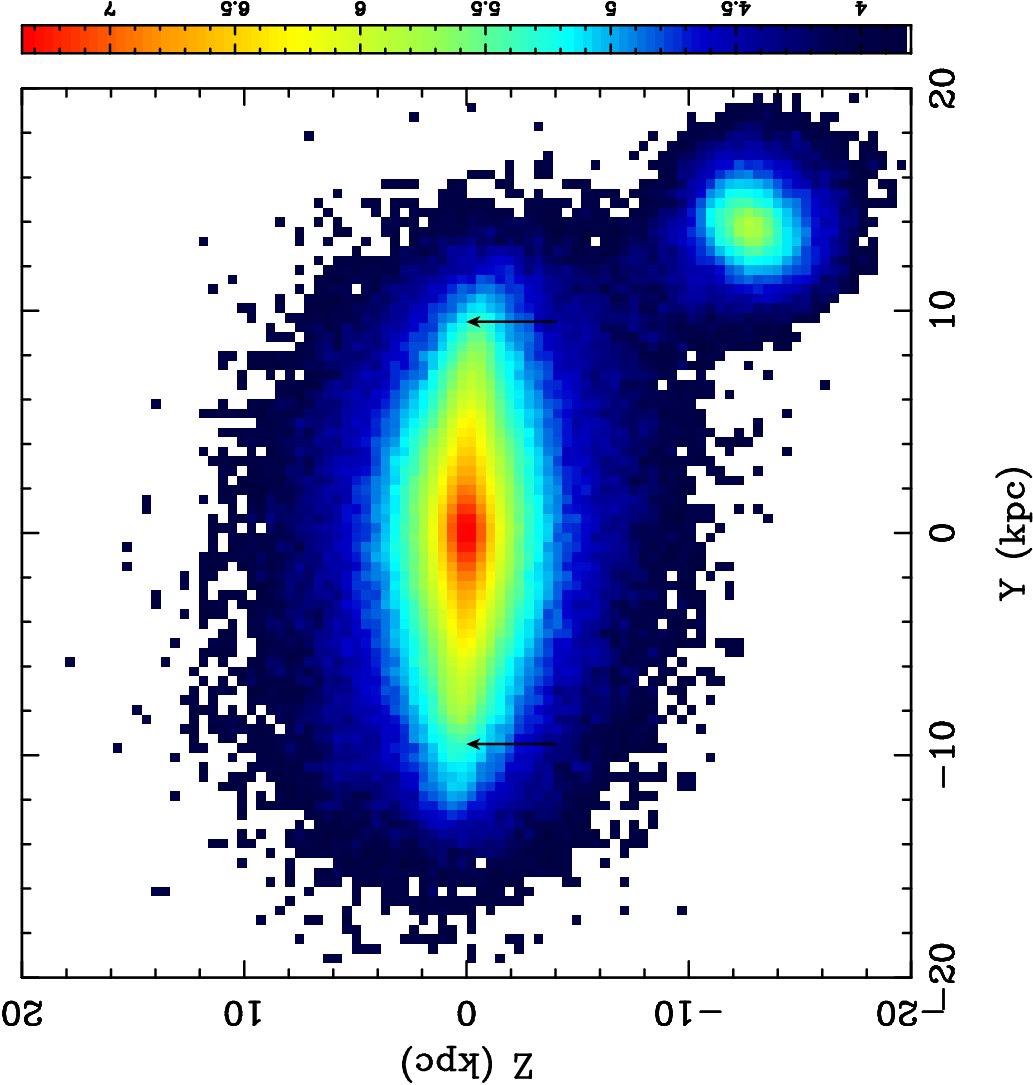}}
\caption{Stellar density projections of the disc (left: face-on;
right: edge-on).  A 40$\times$40$\times$~kpc region is shown in both panels.
The circle in the left panel and the arrows in the right panel indicate the 
position of the break radius (see Sec.~\ref{sec.charac}).
\label{disk}}
\end{figure*}

Figure~\ref{disk} shows a density map of the stellar disc from two
different perspectives (face-on and edge-on). 
The disc extends to $\sim$20~kpc, with an exponential scale-length 
(measured in the V-band) of $\sim$3.2~kpc, consistent with values
found for late-type spirals such as the Milky Way (e.g.
Juri\'c et~al. 2008 suggest a scale-length for the Milky Way's disc
of $\sim$2.6~kpc).

While the stellar disc extends to $\sim$20~kpc, the bulk of our 
analysis is restricted to stars within a radius of 15~kpc.
This is mainly due to the ``law-of-diminishing-returns'' in the outskirts
where the number density of stellar particles decreases dramatically
and the root-mean-square (RMS) dispersion of relevant
azimuthally-averaged quantities begins to increase rapidly.
Furthermore, considering only particles
inside this 15~kpc radius means we also reduce the possible 
contamination from co-spatial
stellar halo particles. 

We end by noting that all ``error bars'' plotted throughout the paper are
derived from the RMS associated with the mean values from eight
arbitrary octants of the disc.

\section{Characteristics of the Disc Break}
\label{sec.charac}
\begin{figure*}
\resizebox{\textwidth}{!}{\includegraphics[angle=-90]{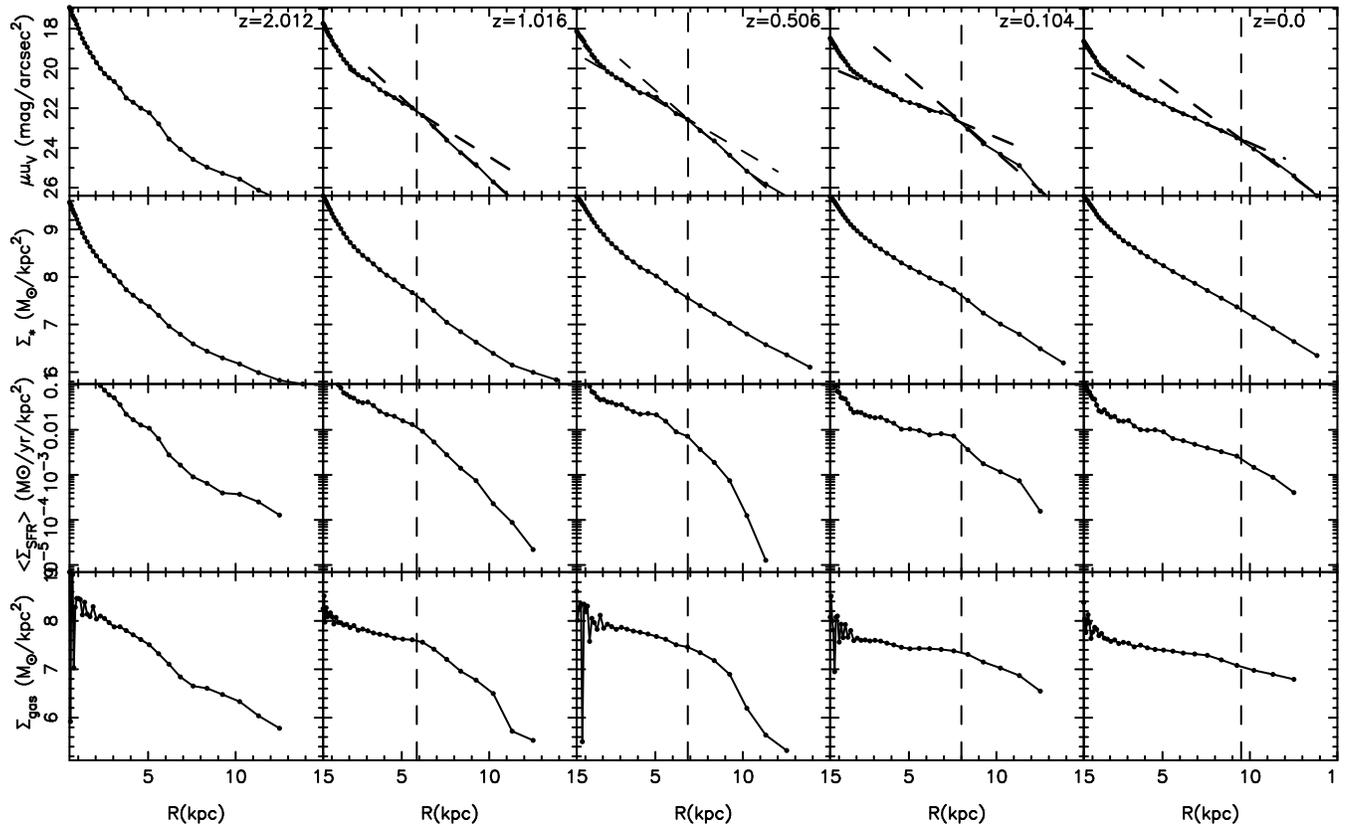}}
\caption{Evolution with redshift -- as indicated in the inset to 
each panel -- of the V-band surface brightness profile (first row), total
stellar surface density profile (second row), 
star formation rate surface density profile (third row), and gas 
surface density profile (fourth row), all on logarithmic scales.
The profiles were derived using all stars with
r$<$15~kpc and $|$z$|<$3~kpc. Dashed lines represent fits to the 
inner and outer exponentials in the V-band
surface brightness profile. The break radii correspond to the 
intersection point of these two exponentials, and are noted by the 
vertical dashed lines in each panel.  
\label{bigplot}}
\end{figure*}

The first two rows of Figure~\ref{bigplot} show the evolution of the
azimuthally-averaged surface brightness profile in the V$-$band and
the total stellar surface density profile, respectively, at five 
different epochs from $z$=2 to $z$=0.
It can be seen that the disc has already developed a break by $z \sim
1$ in the surface brightness profile and maintained its presence
through to $z$=0.
We fit simple functional forms
($\mu(R)=\mu_0+1.086\times R/h$) to the inner and outer exponentials
and consider that the break occurs at the position where the
functions intersect.
At redshift $z=0$, the break radius is $R_{\rm br}$$\approx$9.5~kpc
(i.e., $\sim$2.9 disc scale-lengths).
The break intensity -- measured as the
angle between the inner and outer exponential profiles -- changes with
redshift, but not in any obvious monotonic/systematic manner.
It can also be
seen that the break is much shallower in the total stellar mass surface
density distribution.  In fact, in most snapshots it is almost non-existent,
a point to which we return in Section~\ref{sec:densityprofile}.

\begin{figure*}
\centering
\psfig{figure=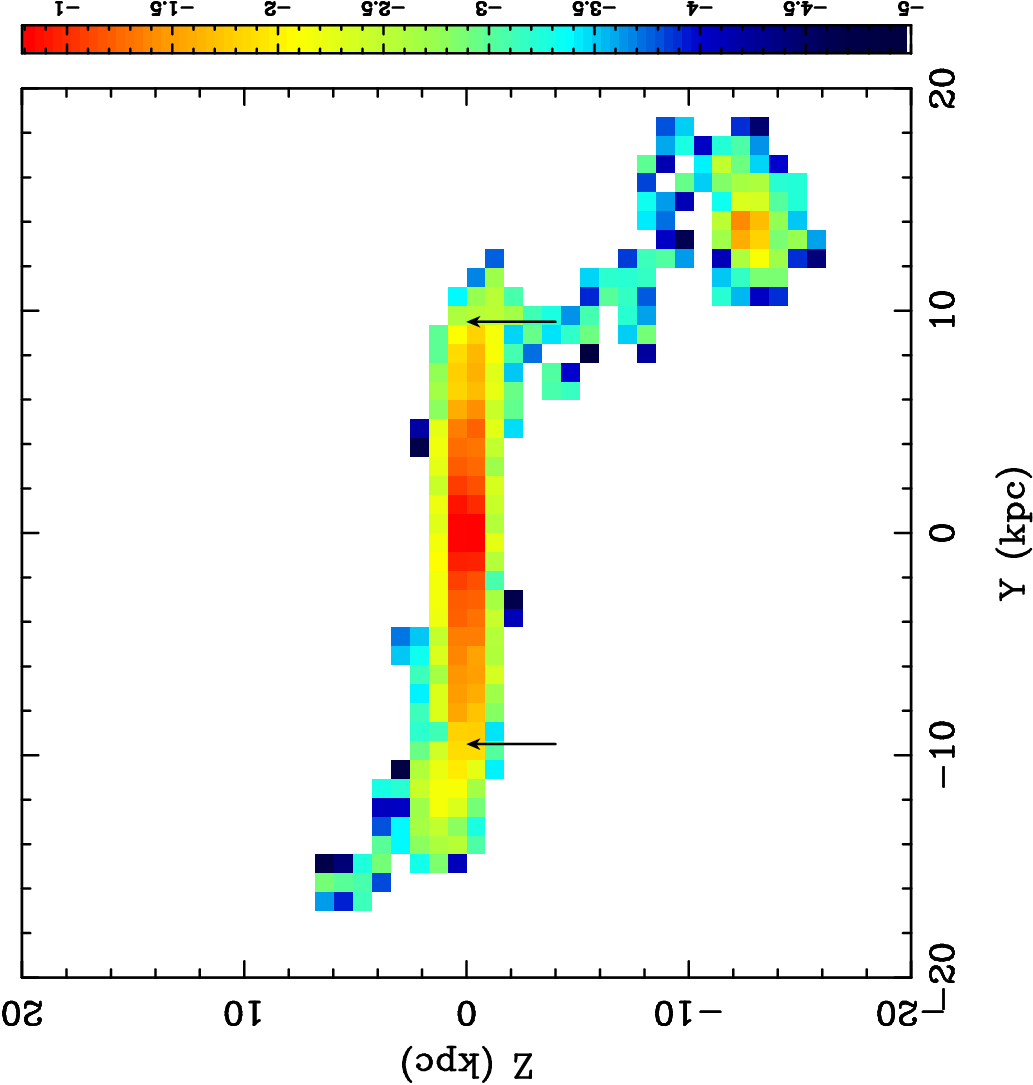,angle=-90,width=6cm}
\psfig{figure=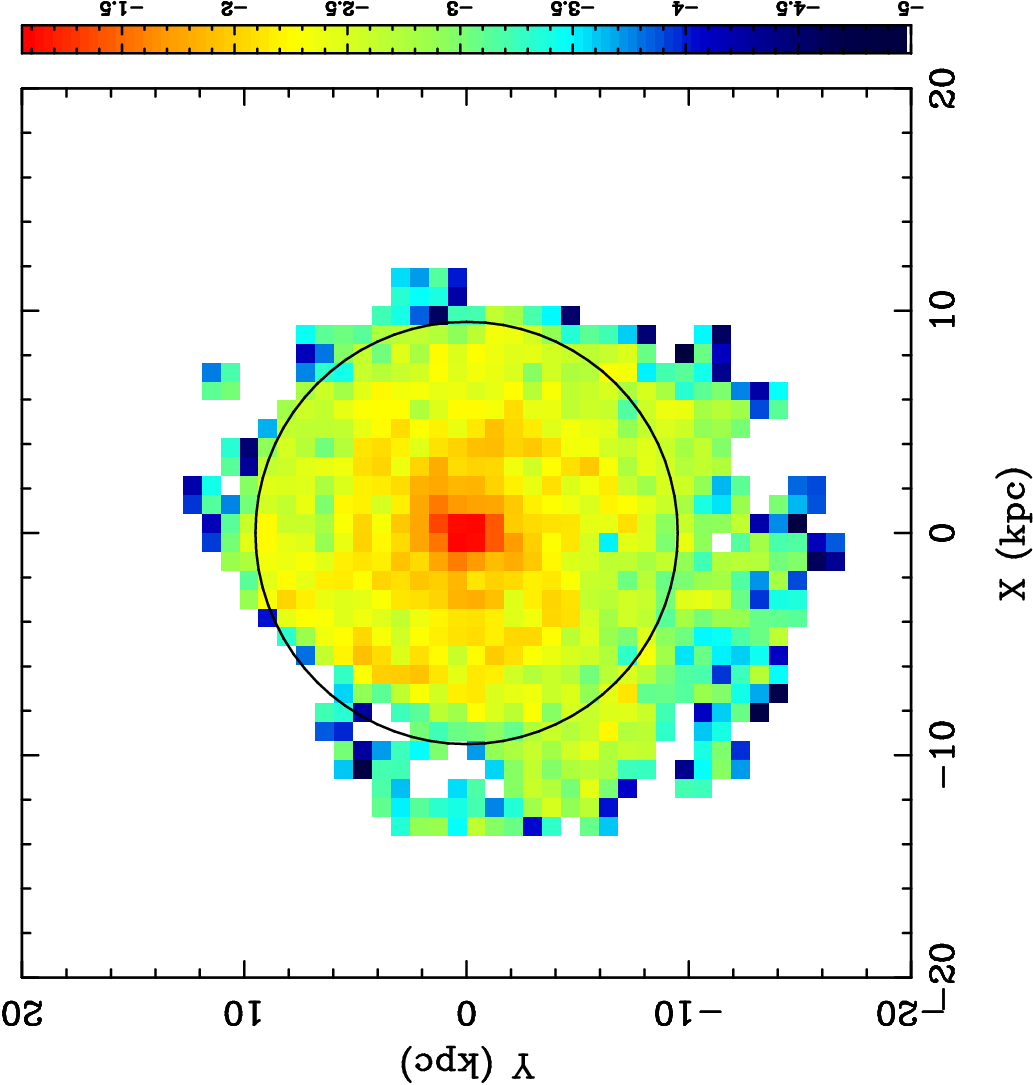,angle=-90,width=6cm}
\caption{Edge-on and face-on 2-dimensional projections of the present-day
integrated (over the last 100~Myr) SFR for our cosmological
disc. For the face-on projection, only the stars at a distance of 5~kpc 
from the plane are included,  to avoid contamination from the companion. The total
line-of-sight in the edge-on projection is 40 kpc.The vertical arrows (edge-on) and circle (face-on) correspond to
the position of the break radius in the V-band surface brightness 
distribution.\label{disk.sfr}}
\end{figure*}

Figure~\ref{disk.sfr} shows spatially-resolved 2-dimensional
star formation rate (SFR) maps of our simulation at $z$=0.
The break radius is denoted by the vertical arrows (edge-on view)
and circle (face-on view); an immediate point of interest is that
there remains measurable star formation outside the break radius.
This star formation is not axisymmetric, with star formation 
extending beyond $R_{\rm br}$ on one side, in particular. Such star
formation in the outer disc (beyond the break radius) has been
observed in many galaxies \citep[][]{Fer98PhD, Fer98, Thil05, Thil07, 
GdP07}.  
Before assessing the physical origins driving the formation of
the break in the
surface brightness profile, we first review the main observational
characteristics, to determine if our light profile is a fair 
representation
of those observed in real galaxies.

\subsection{Dependence of the break upon distance from the mid-plane}

\begin{figure}
\resizebox{0.4\textwidth}{!}{\includegraphics[angle=-90]{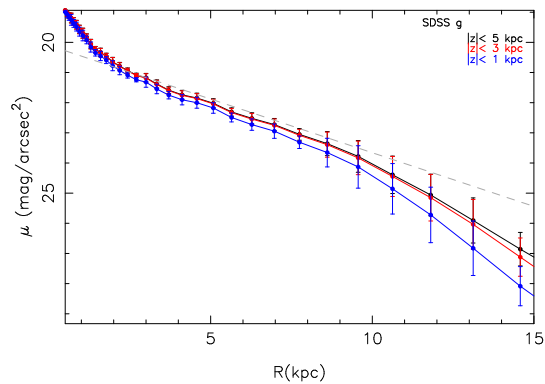}}
\caption{Azimuthally-averaged surface brightness profiles in the SDSS g-band,
integrating particles within 1, 3, and 5~kpc from the mid-plane, as indicated
in the inset. Error bars represent the RMS dispersion between the mean 
values found in eight arbitrary octants of the disc.
\label{scaleheight}}
\end{figure}

In the case of NGC~4244, 
\citet{dJ07} show that
the break occurs at the same radius independent of height above the mid-plane.
They further demonstrate that the intensity of the break decreases with
height above the plane.
Figure~\ref{scaleheight} shows the surface brightness profiles derived
from our simulation, in the 
SDSS g-band,
for samples of stellar particles at different distances from the plane.
The empirical characteristics seen by \citet{dJ07} are similarly 
seen in the simulation -- specifically, the 
break radius position does not change 
with distance from the mid-plane, but it gets shallower.

\subsection{Dependence of the profile upon photometric band}
\label{break.band}

Different photometric bands are sensitive to different stellar populations.
While the main contributors to the luminosity in the blue bands are the young 
stars,\footnote{Other hot stellar populations, including 
blue horizontal branch stars and blue stragglers, may provide
an additional contribution in some systems (see, e.g., 
Rose et al. 1985\nocite{Ros85}; Maraston \& Thomas 2000\nocite{MT00}; 
Trager et al. 2005\nocite{Tra05})} this contribution decreases
for the redder bands.
Interestingly, \citet{Bak08} have found recently
that the position of the break is independent of the photometric bandpass
employed to obtain the 
surface brightness profile, although break is shallower in the redder
bandpasses.
These results suggest that the break is due mainly to a truncation in the 
profile of the young stars and that the profile of the old stars should 
not show the same departure from a pure exponential.

\begin{figure}
\resizebox{0.5\textwidth}{!}{\includegraphics[angle=-90]{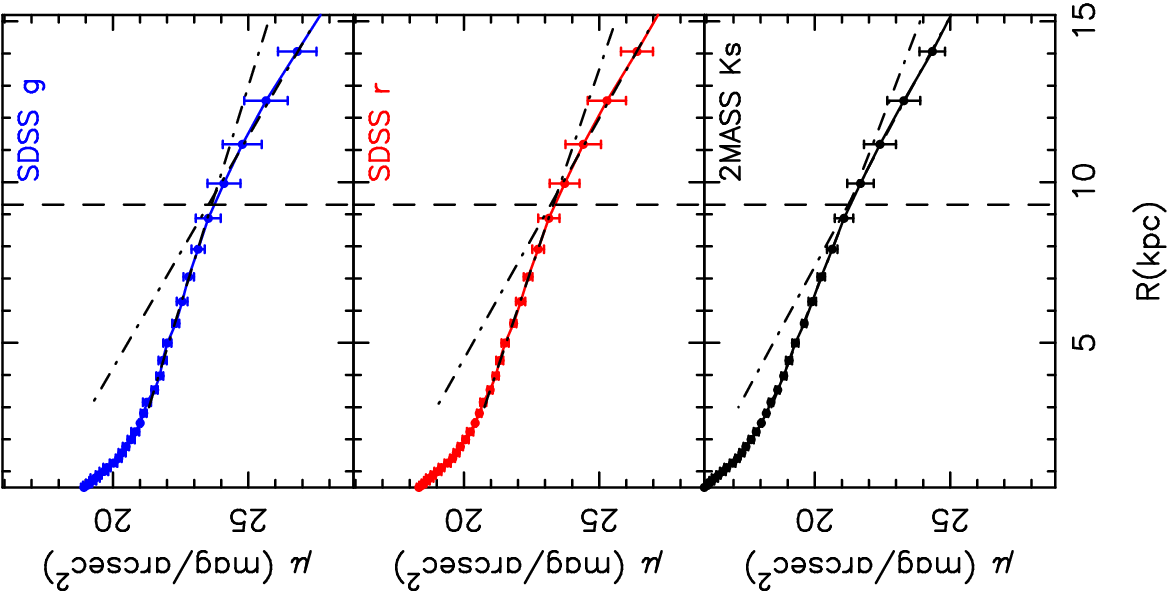}}
\caption{Surface brightness profiles measured in three different bands, 
as indicated in the inset to each panel.
Dashed vertical lines show the position of the break radius, 
while the dashed-dotted lines represent the fits to the 
inner and outer exponentials. \label{surf.bands}}
\end{figure}

Figure~\ref{surf.bands} shows the surface brightness profiles of the
stellar disc in the same two photometric bands (g and r)
used in the empirical
study of Bakos et~al. (2008), in addition to the redder K-band.
In our simulation, the
break appears at the same position ($\sim$9.5~kpc) but, in agreement
with the observations, is also shallower in the redder band (Ks) than
it is in the bluer one (g).  Furthermore, Bakos et~al. (2008)
transform the surface brightness profiles into total stellar mass
density profiles, using mass-to-light ratios inferred from the
colours, following the prescriptions of \citet{Bell03}. This process
is admittedly not free of uncertainties, but, at face value, their
results suggest that the mass density profile does not show a break,
in agreement with that found for our simulated disc (see
Fig.~\ref{bigplot}).

\begin{figure}
\centering
\resizebox{0.4\textwidth}{!}{\includegraphics[angle=-90]{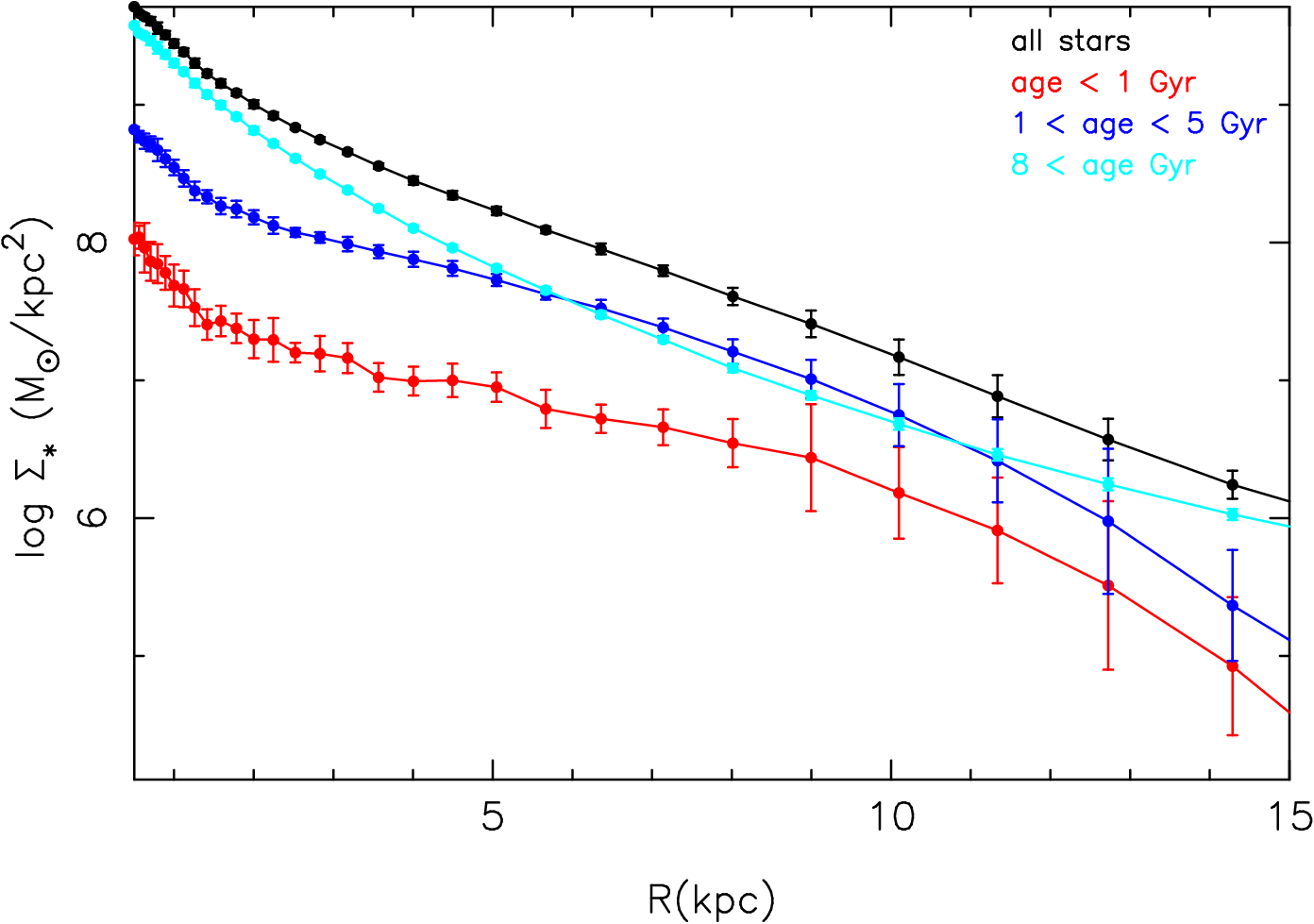}}
\caption{Stellar surface density profiles for stars of different ages,
as noted in the inset.
The intensity of the break decreases with increasing age of the stellar
population under consideration, with old stars showing an ``upbend'' 
at large radii.  The total stellar surface density profile is 
consistent with a pure exponential out to $\sim$5 exponential scale-lengths.\label{dens.age}}
\end{figure}

Figure~\ref{dens.age} shows the stellar mass surface density
distribution in our simulated disc for stars of different ages.  While
the position of the break is the same for those stellar populations
showing a break in the stellar density distribution, not all stars
show this truncated density profile. In particular, stars older than
$\sim$8~Gyr show, if anything, a slightly ``upbending'' profile.  This
latter result agrees with a number of observational studies:
\citet{Dav03} studied the stellar populations in the outskirts of
NGC~2403 and M33, showing that while the number of young main-sequence
stars present a ``downbend'' with radius, the number density of the
more evolved stars in the red- and asymptotic- giant branch phases do
not; Similar results were also obtained by \citet{Gall04} for M33.
Conversely, for the case of NGC~4244, \citet{dJ07} claim a sharp break
even when only red giant stars (old stars for a constant star
formation history) are considered.  This issue clearly needs resolving
with a larger sample of galaxies.

\subsection{Colour profiles}

Recently, \citet{Bak08}, for the local universe, and \citet{Azz08}, out
to redshift z$\sim$1.1 (see also \citet{Jan00}) have found that the
colour gradients of galaxies with Type-II truncations are
different from those with Type-I or pure exponential profiles (recalling
the nomenclature introduced in Section~1). In
particular, galaxies showing Type-II truncations have a ``U-shaped''
colour profile, with the minimum (bluest color) at the position of the
break.

\begin{figure*}
\resizebox{0.75\textwidth}{!}{\includegraphics[angle=-90]{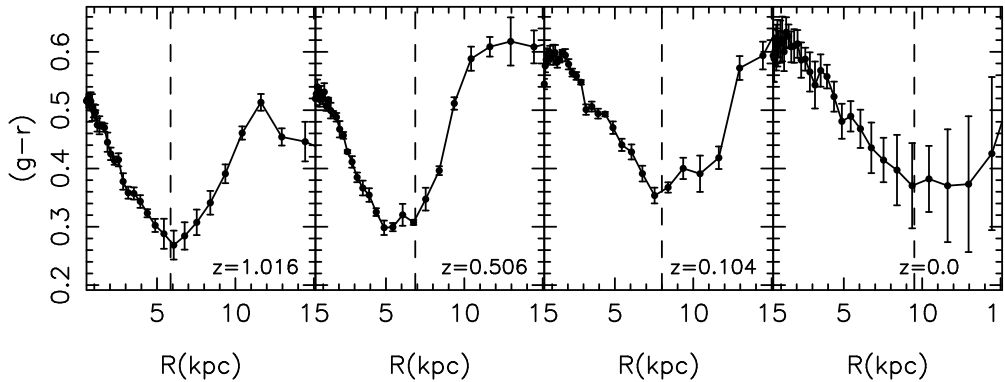}}
\caption{Colour profiles for several outputs of our simulation from $z=1$
to $z=0$.
Dashed lines in each panel indicate the position of $R_{\rm br}$ in the V-band
surface brightness 
profile (recall, Figure~\ref{bigplot}). Error bars reflect the RMS dispersion
of the mean azimuthally-averaged colour in randomly-selected octants of the 
disc.
The colour profiles show a minimum at the break radius, in agreement with
observation.\label{colour-prof} }
\end{figure*}

Figure~\ref{colour-prof} shows our predicted colour profile at four
different epochs from $z$=1 to $z$=0.  In agreement with the 
observations, our simulated disc also shows a U-shaped profile with
the bluest colours near the position of the break.

\subsection{Age and metallicity profiles}
\label{ageprofile}

\begin{figure}
\resizebox{0.5\textwidth}{!}{\includegraphics[angle=-90]{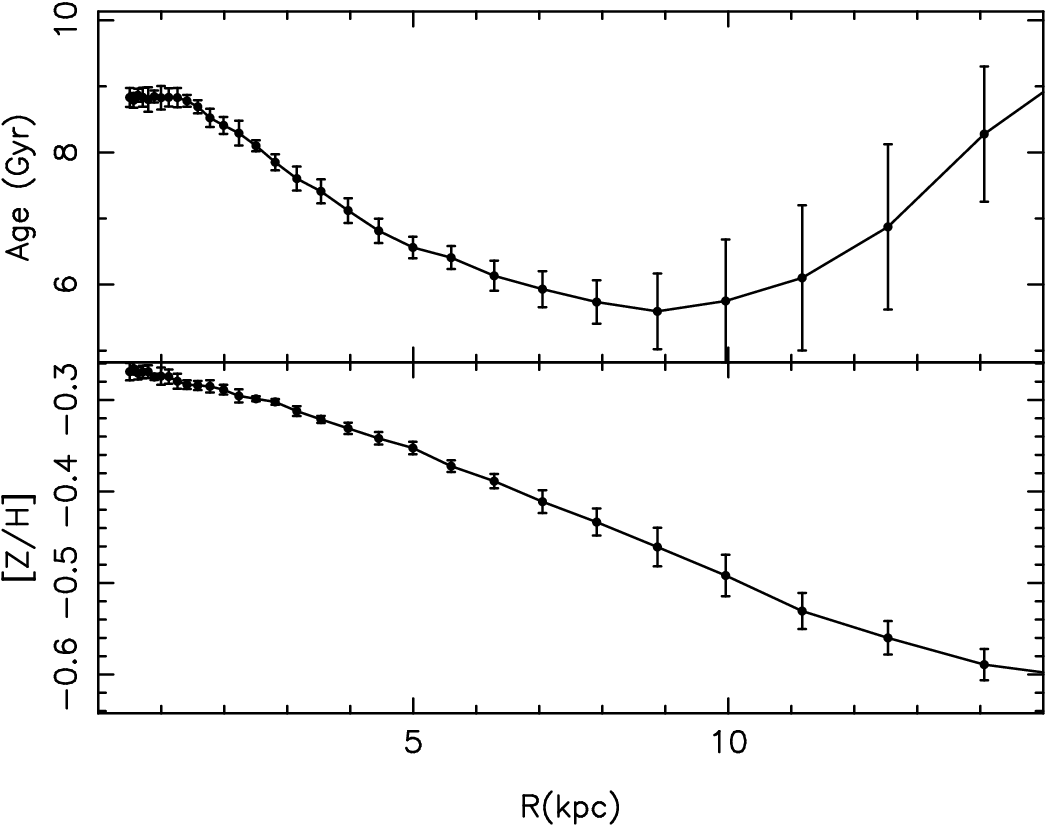}}
\caption{Mass-weighted age and metallicity gradients for our simulated disc
at $z$=0.\label{profiles}}
\end{figure}

Colours are difficult to interpret in terms of simple stellar populations
due to the well-known ``age-metallicity degeneracy'' \citep{Wor94};
in principle, the colour profile could be an effect of age,
metallicity, or a combination of both.  Figure~\ref{profiles} shows the
mass-weighted stellar age and metallicity gradient for the final
output of our simulation. It can be seen the age profile presents a
minimum at roughly the position of the break. However, the metallicity
profile is remarkably smooth all the way out $\sim$15~kpc, without any
visible change at the break radius.  Therefore, the U-shape of
the colour profiles seen in previous section is due to an age effect, and
not to a metallicity effect.

\subsection{Evolution of the break with redshift}

Several studies have now analysed the empirical 
evolution of the break radius with
redshift \citep{Per04, TP, Azz08}.  Using a sample in excess
of 200 galaxies, for example, Azzollini et al. suggest that,
for a given stellar mass, the
radial position of the break has increased with cosmic time by a
factor of 1.3$\pm$0.1 between $z\sim 1$ and $\sim$0. They also found
that, in the same period of time, the evolution of the surface
brightness level in the rest-frame B-band at the break radius
($\mu_{\rm br}$) has decreased by 3.3$\pm$0.2~mag~arcsec$^{-2}$.

\begin{figure*}
\resizebox{0.9\textwidth}{!}
{\includegraphics[angle=-90]{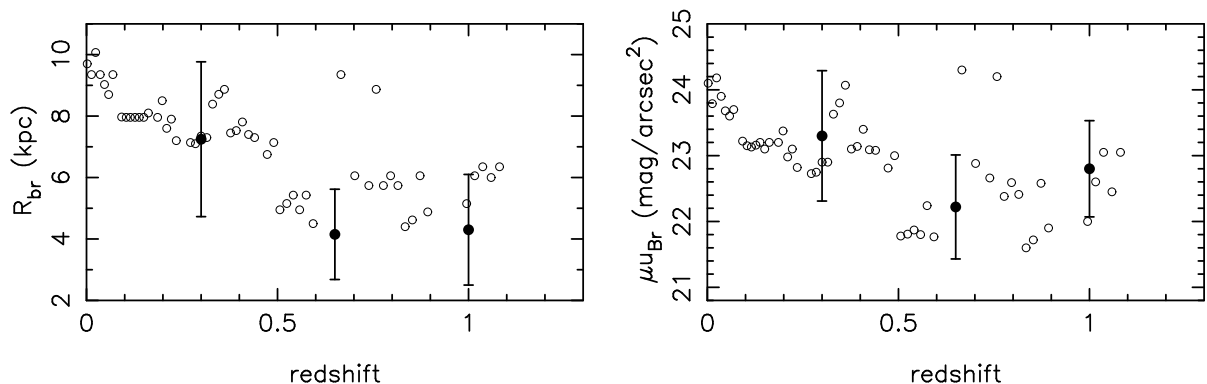}}
\caption{Predicted evolution with redshift of $R_{\rm br}$ and 
$\mu_{\rm br}$ (open circles) compared
to the observed values from Azzolini et al. (2008) (filled circles 
with error bars). The error bars show the RMS-dispersion 
of the measured values for galaxies in a given 
redshift- and mass-bin.\label{break.z}}
\end{figure*}

Because the mass of our simulated galaxy evolves with redshift, we
cannot compare directly the evolution of these parameters for a galaxy 
with a given mass. For this 
reason we 
re-analysed these data (kindly provided by the authors).  
 We separated the observational data into three redshift bins
and fitted, at each redshift, 
a linear relation between $R_{\rm br}$,
$\mu_{\rm br}$, and the mass of the galaxies, in an identical fashion to that 
done by Azzollini et~al. (2008). Using the linear fit, we measured  
the predicted values of
$R_{\rm br}$ and $\mu_{\rm br}$ for a galaxy with the same mass as
our disc at each of
the different redshift bins (note that the mass is different at 
each redshift) and their associated RMS dispersions.  
Figure~\ref{break.z} shows
the predicted evolution of $R_{\rm br}$ and $\mu_{\rm br}$ in the
B-band for our simulated disc, compared to the observed data.  
It can be seen 
that the evolution of these parameters within our simulated
disc are very similar to that which is observed.

\subsection{Age and metallicity distribution in the outskirts}

\begin{figure}
\resizebox{0.4\textwidth}{!}{\includegraphics[angle=-90]{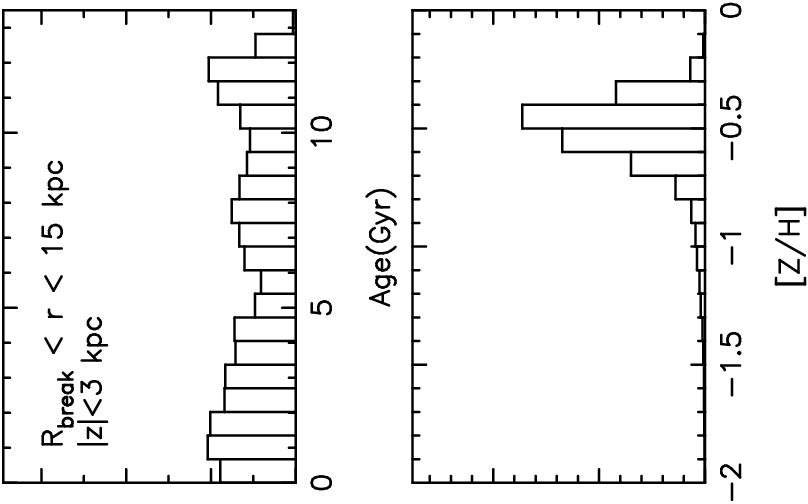}}
\caption{Age and metallicity distributions for the stars with radius 
R$_{\rm br}$$<$$r$$<$15~kpc and $|$z$|$ $<$ 3~kpc.\label{age.dis}}
\end{figure}

Studies of stellar population in the outskirts of discs have found
stellar populations which are of old-to-intermediate ages, with
metallicity distributions peaking at relatively high metallicities
([Fe/H]$\sim -$ 0.7) (e.g., Ferguson \& Johnson 2001\nocite{FJ01}
(M31); Davidge 2003\nocite{Dav03} (NGC~2403, M33); Galleti et
al. 2004\nocite{Gall04} (M33)).  It has been suggested that these
properties are not expected in CDM models (e.g., Ferguson \& Johnson
2001), where the disc grows inside-out and at relatively recent epochs
($z$$\simlt$1).  Figure~\ref{age.dis} shows the age and metallicity
distributions of stars with R$_{\rm br} < r < 15$~kpc in our simulated
disc.  There is a very large scatter in the ages of the stars and a
significant fraction of the stellar population has ages in excess of
$\sim$8~Gyr.  The peak of the metallicity distribution is near
[Fe/H]$\sim$$-0.5$ -- i.e., even larger metallicities than that
normally observed in the outskirts of galaxies.  However, it should be
noted that the ``observed'' metallicities are inferred via the use of
colour-magnitude diagrams. When dealing with photometric
metallicities, one must bear in mind that the underlying age
distribution may affect the derived metallicity distribution. All of
the studies cited above obtained their metallicites by comparing with
the sequences of old Milky Way globular clusters. If a scatter in age
is allowed, with a larger fraction of young stars (as we obtain in our
simulation), the metallicity distribution will be skewed towards
higher values and more closely resemble our derived distribution.  We
note that our metallicity distribution has a metal-poor tail, as
reported in some empirical studies (e.g., Galleti et~al. 2004).

\section{Break Formation}

As foreshadowed in Section~1, there is no clear consensus regarding
the underlying physical mechanism responsible for the surface
brightness profile breaks observed both locally and at high-redshift.
Broadly, there are two categories of theories -- those related with
the angular momentum distribution of the stars and those related to
star formation thresholds.

In our simulated disc, the break is seen in the light profile, but not in the
stellar mass density profile.  This indicates (or at least
suggests) that the break in our disc is
likely related to a change in the stellar population properties
and, therefore, with a star formation threshold.  The third and fourth rows
of Figure~\ref{bigplot} show the SFR density (with the SFR averaged over
the last 1~Gyr) and gas surface density 
profiles, respectively.
Examining 
Figure~\ref{bigplot}, we can see that
the gas surface density at the break radius is not
the same at all redshifts.  For example, at $z=2.01$ there is no clear
break in the surface brightness profile, despite the gas surface density
reaching values significantly below the star formation threshold.
This would seem to contradict
the aforementioned 
star formation threshold scenario as the driver behind the break.  In
\citet{Rok08a}, the reason for the break was a sudden decrease in the
gas density, but also, in agreement with our results, the density of
the gas at that radius was higher than the star formation threshold
imposed in their simulations.  In their work, the decrease in the
stellar density was due to an angular momentum limitation.  By
construction, in their simulations, the angular momentum is directly
proportional to the radius, which means that the high angular momentum
material will take longer to cool.  Therefore, the angular momentum
determines the maximum extent of the gaseous disc and of the 
star-forming disc.  In fully cosmological simulations, however, the
accretion of matter is not as regular as the classical model of
spherical shells collapsing from increasingly large
radii. Particularly, at high-redshift, before the galaxy is large enough
to shock the accreting gas, accretion of gas occurs along filaments,
as well as in mergers and clumps. The growth of our simulated disc is thus
significantly more complicated than the disc in Ro{\v s}kar et~al.'s idealised
models. Although our disc does grow ``inside-out'', it does so in a less
regular manner.  In our simulations, in fact, the gaseous disc extends
beyond the stellar disc and, while in {\it some} timesteps the position of
the break coincides with a change of slope in the surface gas density
profile, this is not true at {\it all} timesteps.

\begin{figure*}
\centering
\psfig{figure=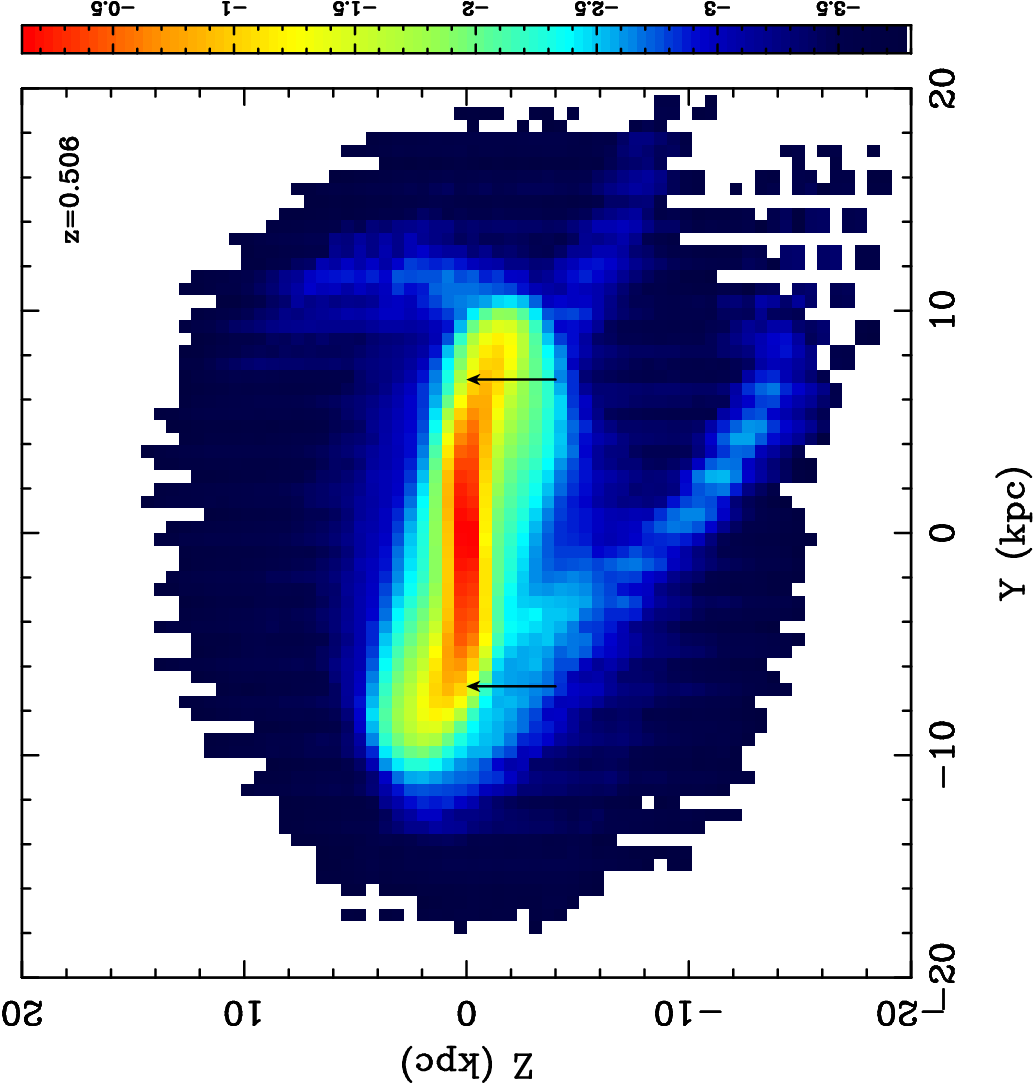,angle=-90,width=4.2cm}
\psfig{figure=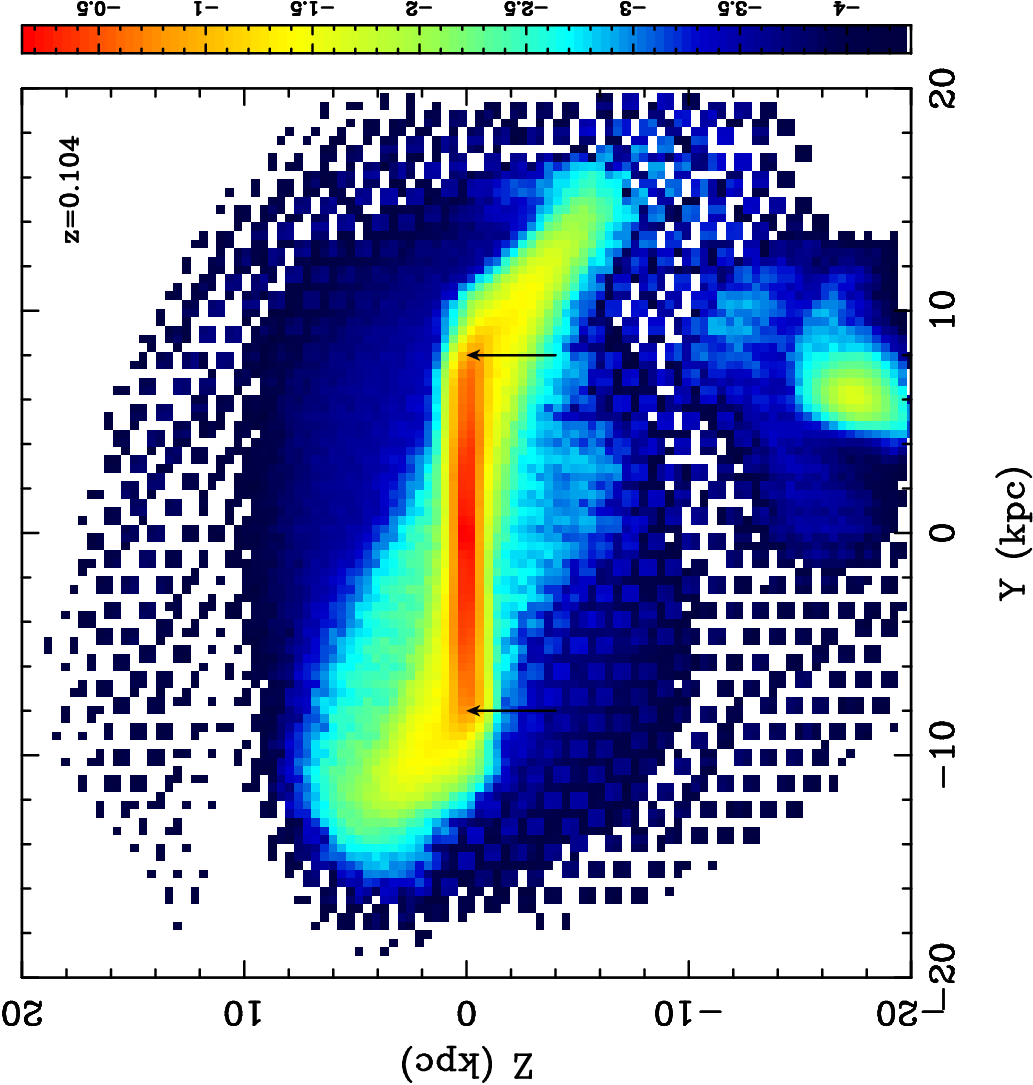,angle=-90,width=4.2cm}
\psfig{figure=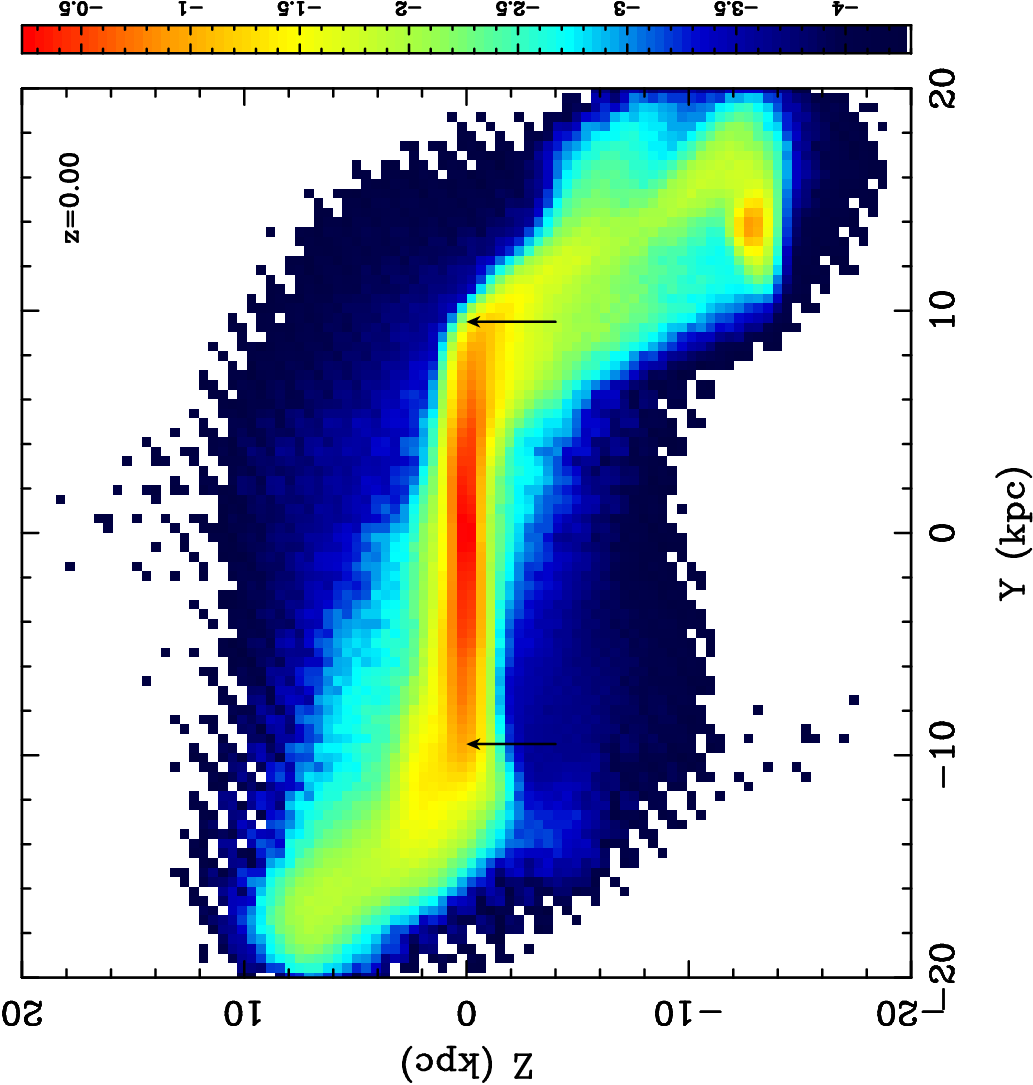,angle=-90,width=4.2cm}
\caption{Edge-on projections of the logarithmic hydrogen density in H cm$^{-3}$, 
at three different redshifts (noted in the inset to each panel).
Vertical arrows indicate the position of the break in the associated 
V-band stellar surface brightness profiles.\label{gas.map}}
\end{figure*}

\begin{figure}
\centering
\psfig{figure=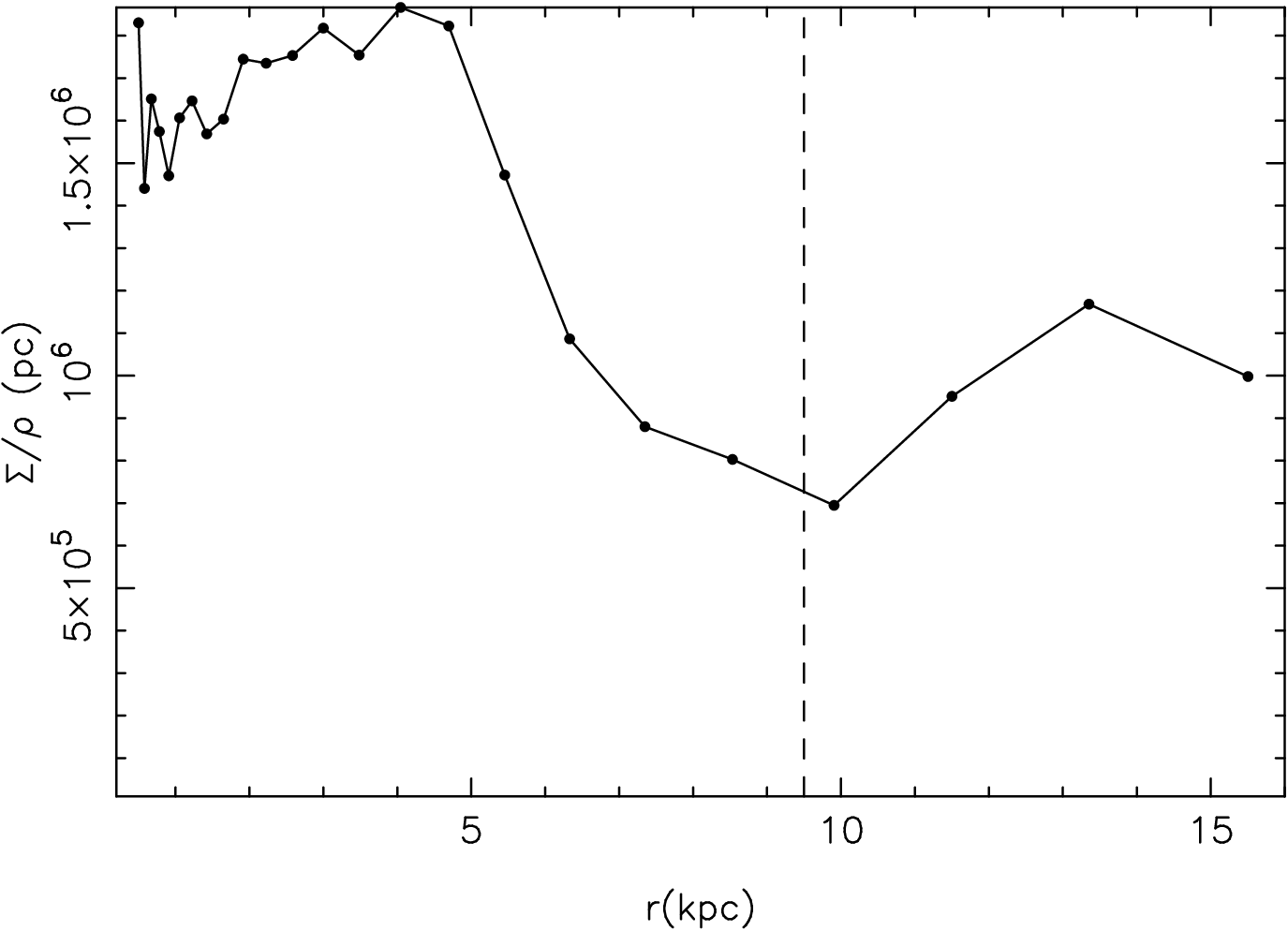,angle=-90,width=6cm}
\caption{Ratio between the surface mass density and the
volume density, as a function of galactocentric radius at redshift $z$=0,
in our simulated disc galaxy. The position of the break in the surface
brightness profile is indicated with a dashed line.\label{densitiesratio}}
\end{figure}

\citet{Fer98} proposed that the breaks in the star formation density
could be due to an intrinsic correlation between azimuthally-averaged
star formation rate and gas volume density, combined with a vertical
flaring of the disc or a warp, in such a way that the transformation
between gas surface density and volume density varies with
galactocentric radius \citep{Mad74}.  We plot, in
Figure~\ref{densitiesratio}, the ratio between the surface density and
the volume density of the gas in our simulated galaxy. It can be seen
that, indeed, there is a change in the slope of this ratio at
(approximately) the break radius.

Drops in the surface density of the gas at the onset of a warp
have been observed in several galaxies \citep{Jozsa07,
GR02}. In fact, a possible connection between truncated profiles and
gaseous warps has been suggested several times in the literature
\citep{vdK07}.  Figure~\ref{gas.map} shows the edge-on projection of the
gas density distribution of our disc at three different redshifts,
matching those of Figure~\ref{bigplot}.  The break radius coincides with
the onset of a warp in the gaseous distribution in each of these three
time-steps.  Therefore, the
decrease in the gaseous volume density due to this flaring seems to be
a fundamental mechanism responsible for the break seen in the 
average star formation rate and, hence, the break in the light profile.

The coincidence of the initiation of the warp with the break 
radius at different timesteps is an encouraging piece of evidence
in support of this suggestion.  Having said that, this is but one 
simulation; a definitive statement awaits the analysis of a systemic
suite of 20-30 comparable simulations, sampling a range of
environments, assembly histories, and dynamical masses -- such
an ambitious program is currently underway within our group.

\section{Radial Migration}
\label{sec:migration}

\begin{figure*}
\resizebox{0.3\textwidth}{!}{\includegraphics[angle=-90]{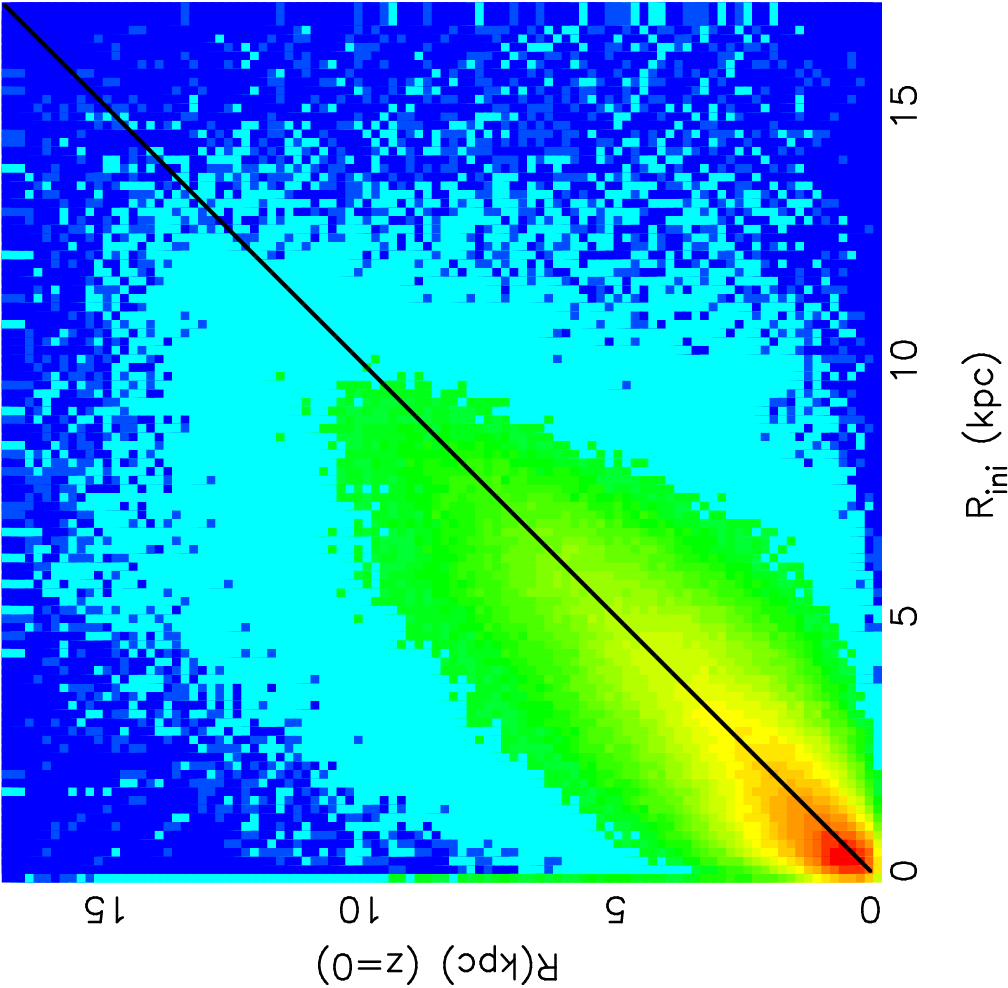}}
\resizebox{0.5\textwidth}{!}{\includegraphics[angle=-90]{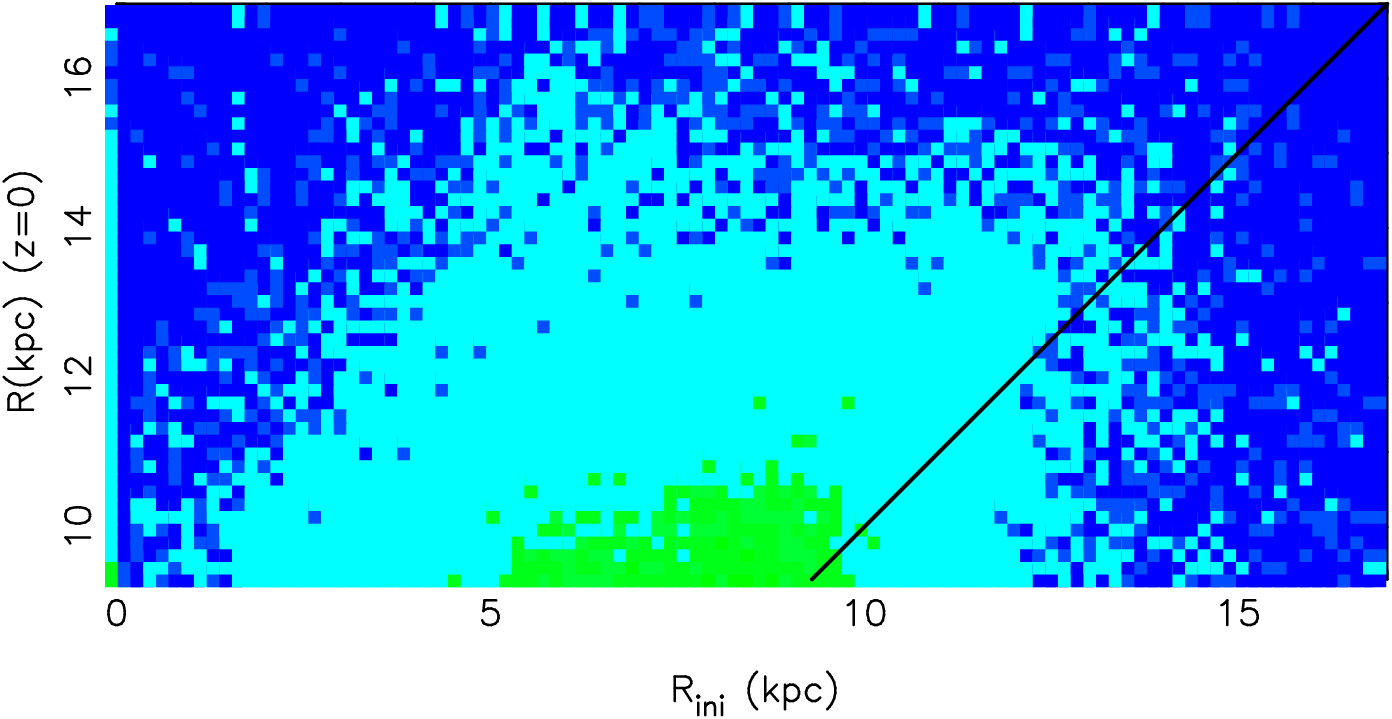}}
\caption{Final stellar galactocentric radius versus formation radius.
The different panels correspond to 
different scalings on the abscissa.\label{migration}}
\end{figure*}

Figure~\ref{migration} compares the final (at $z$=0) galactocentric
radii for stars in our simulated disc 
(R$_{\rm final}$) with the radius at which they were formed
(R$_{\rm initial}$).  The right panel shows the same distribution, 
but only
for those stars outside the break radius.  Using disc stars 
with $R_{\rm br} < r <$15~kpc, we find that $\sim$57\% of the stars
formed inside the break
radius, $\sim$21\% formed {\it in situ}, and $\sim$22\% formed at
$r>$15~kpc or $|$z$|> 3$~kpc, including those formed elsewhere in
external satellites -- i.e., more than half of the stars currently
in the outskirts of the disc formed at lower radii.

The mean radial distance traversed by the disc star 
particles (all those
with 3$<r<15$~kpc and 
$|$z$|<$ 3~kpc, discounting those formed in satellites for this 
calculation)
is $<$$|$$R_{\rm final}$$-$$R_{\rm 
initial}$$|$$>$=1.7~kpc, while for those with
$R_{\rm br}<r<$15~kpc, this mean radial 
traversal distance is 3.4~kpc.
It is clear that in our simulation there is considerable radial migration of
stars towards the external parts. 
The importance of such radial migration and mixing within the
thin disc has been the subject of several important studies
\citep[e.g.][]{SS53, BW67, Fuchs01,SB02, Rok08a, Hay08}.

It has been known for a number of years that scattering by 
spiral structure and molecular clouds can gradually heat
stellar discs, moving stars towards more inclined and 
eccentric orbits and changing the overall angular momentum 
distribution of the disc. Minor mergers and accretion 
of satellities can also produce heating  
(e.g., Velazquez \& White 1999, amongst many others).
Stars that are on eccentric orbits 
are at different radii at different phases of their orbits and,
therefore, tend to naturally produce some radial mixing.  However, the 
radial excursions due to this mechanism are not sufficient to 
explain the flat age-metallicity relation in the solar 
neighbourhood (Sellwood \& Binney 2002).
Typical radial variations for a population of stars with a radial 
velocity dispersion $\sigma_r$ are $\Delta$R$\sim \sqrt{2} \sigma_r/k$, 
where $k$ is the epyciclic frequency; for the old stars in the 
Milky Way the maximum value for the excursions near the Sun is 
$\sim$1-1.5~kpc.

Another proposed mechanism for radial migration was 
called ``churning'' by Sellwood \& Binney (2002), and
is due to scattering of stars across co-rotation resonance
by spiral waves  \citep[][]{SB02, SP02, SK02, Rok08a},
resulting in a change in the orbit centres, but not the eccentricity.
Such churning causes little increase in random motions (or
heating) because it preserves the overall distribution of angular
momentum (see Sellwood \& Binney 2002 for details).
In the work of \citet{Rok08a}, it was this churning mechanism that
was predominantly responsible for the radial migration.

Our cosmological disc (as all hydrodynamical cosmological discs to
date have been) is significantly hotter than that of the Milky Way
(Gibson et~al. 2008); as such, radial excursion of stars in eccentric
orbits is no doubt contributing to (and perhaps dominating) the
mixing. In fact, the epyciclic radius of the outer disc particles is
3.3~kpc, compatible with the mean calculated $\Delta$R.  We expect
radial churning to have an important contribution too, although due to
the larger scale-heigh of our disc, this contribution is certainly
less important than in \citet{Rok08a}.  Furthemore, our
cosmological disc is continuously being bombarded by small satellites.
Younger et~al. (2007) showed that merger-driven gas inflows deepen the
central potential and contract the inner profile while, at the same
time, angular momentum is transferred to large radius and causes the
outer disc to expand. This has the net effect of driving radial migration of
the stars towards the external parts.  It is very difficult to
disentangle all the different causes of radial migration in our
cosmological disc,\footnote{Indeed, 
due to the large velocity dispersion
of the stars compared with the Milky Way and to the larger
scale-height, cosmological simulations may not be the most appropriate
for studying these particular mechanisms.}
although it is certainly the case that each of
these mechanisms is playing a role in both our simulated disc
and in real galaxies.  The main heating mechanisms in the disc
presented here and in other cosmological simulations will be the
subject of a future paper (House et~al. 2009, in preparation).

\citet{Rok08b} showed that radial migration due to this
churning mechanism was able to explain the exponential decay of the
surface brightness profile beyond $R_{\rm br}$, the constancy of the
break radius for stars of different ages, and predict age profiles in
agreement with observations \citep{Bak08}.  
The simulation of \citet{Rok08a} was performed under idealised conditions,
with the disc growing {\it by construction} in an inside-out fashion,
with very little star formation outside $R_{\rm br}$ (which itself
increases monotonically with time).  
Hence, the stars beyond the break
in their model {\it must} have migrated there. 
While these controlled, high resolution simulations are {\it necessary} to 
study many of dynamical processes having place on the disc, 
it is also interesting (and important) to study the evolution of the surface
brightness profile
within a fully cosmological context,
where hot and cold modes of gas accretion, and the effects of
mergers and interactions are included. 
We find that the break radius does not
always increase monotonically with time (see, e.g.,
Figure~\ref{break.z}) and the shape of the final profile is influenced
by a variety of external processes. Therefore, in the next
section, we explore the effect of radial migration on the overall
properties of our simulated disc.

\section{Implications of radial migration for disc properties}
\subsection{Density profile}
\label{sec:densityprofile}

\begin{figure}
\resizebox{0.4\textwidth}{!}{\includegraphics[angle=-90]{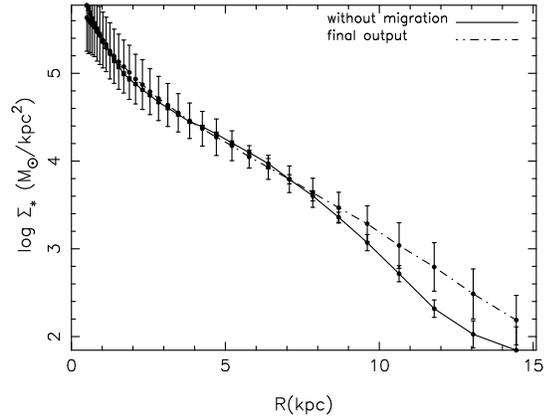}}
\caption{Stellar surface brightness density profile of the disc at $z=0$ (dot-dashed curve) compared
with the  profile the disc would possess in the hypothetical situation
that the redistribution of stellar material by secular processes was absent.
\label{denpro.is}}
\end{figure}

What is the effect of the radial migration in the density profile of
our simulated disc?  Figure~\ref{denpro.is} shows the mass density profile at
$z=0$ compared with the hypothetical density profile that we would
observe {\rm if} the stars had not migrated from their birth place.
The most dramatic changes in the density profile happen in the outer
parts. The movement of the stars tends to weaken the intensity of the break 
that would have otherwise been seen in the stellar mass density profile.

This behaviour is quite different to what is seen in the simulations
where the redistribution of angular momentum is caused by a bar (eg.,
Foyle et al.\ 2008\nocite{Foy08}).  Foyle et~al. found that the outer
profile in their simulations barely changed with time, while the inner
profile became flatter due to the redistribution of material.
Nevertheless, \citet{PT06} distinguish
two types of Type~II profiles -- those associated with the Outer
Lindblad Resonance (OLR), and thus linked with the presence of a bar,
and those not associated with bars, which are normally located at
large radius  (see Erwin et al. 2008 for a more detailed 
discussion). It may well be that two different mechanism are
responsible for these two variants of the Type~II profiles. The
profiles not associated with the OLR are less common in early-type
galaxies than late-type galaxies \citep{PT06} which may indicate that
the presence of an extended gaseous disc are necessary to produce
them.

\begin{figure}
\resizebox{0.4\textwidth}{!}{\includegraphics[angle=-90]{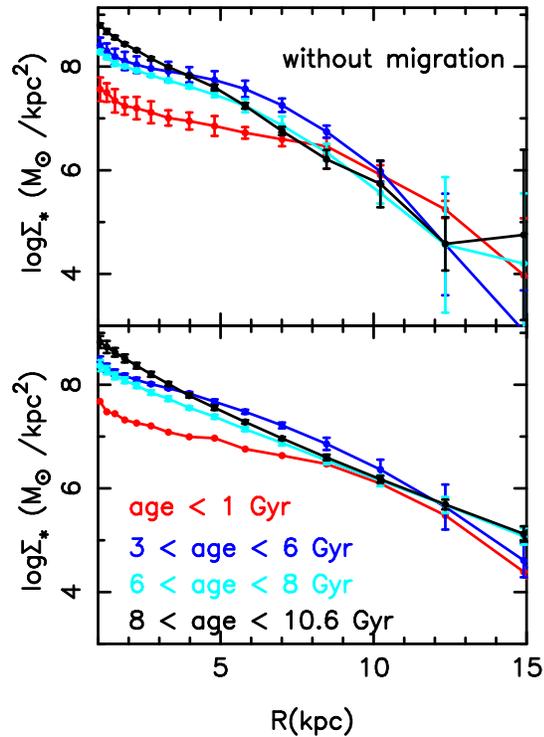}}
\caption{As in Figure~\ref{denpro.is}, but subdividing the stars by 
age, as indicated in the inset to each panel; upper panel: hypothetical
situation of no radial migration; lower panel: as measured in our
cosmological simulation.\label{denpro.ages.is}}
\end{figure}

We showed, in Sec.~\ref{break.band}, that the old stars have a density
profile with an upturn in the outer parts.  Figure~\ref{denpro.ages.is}
demonstrates that this is partially due to the significant effect that
migration has on this outer profile.  In this figure, the stellar
density profile using different age bins is compared with the same
profiles in the hypothetical situation where the stars have not migrated
from the place of their birth. 
It is readily apparent that the surface density profile, in the absence of 
migration, appears truncated in all cases.  This contrasts with the mass 
density profiles that we derived from the final output of our 
simulation, which appear much closer to a pure exponential 
(to at least $\sim$5 exponential scale lengths).

We conclude, therefore, that the exponential mass density profile in
Figure~\ref{bigplot} would appear truncated/broken if there had not been radial
migration.  As old stars have had more time to migrate towards the
external parts, the density profile for the oldest age bin appears
anti-truncated.  This mechanism could also be responsible for the
upbending profiles seen in galaxies where a truncation in the star
formation density does not occur (Type~III profiles, recalling the 
nomenclature of Section~1).  This was also proposed by Younger et~al. 
(2007) who found that in their simulations, the radial 
migration was produced by angular momentum transfer during minor 
mergers; such a scenario receives support from the empirical
evidence associating asymmetries and distortions with anti-truncated
discs (Erwin 2005; Pohlen \& Trujillo 2006).
As mentioned in Section\ref{sec:migration}, several mechanisms are 
certainly operating in our disc, in order to produce the
radial migration of stars toward the external parts, including
this effect of satellite accretion.

\subsection{Stellar populations}

Stars and interstellar gas in galaxies exhibit diverse chemical
element abundance patterns that are shaped by their environment and
formation histories.  A wealth of surveys and satellite missions are
devoted to obtain chemical patterns for individual stars as well as
ages and kinematics in order to derive the star formation history and
merger history of our galaxy (e.g., HIPPARCOS, RAVE, SEGUE, GAIA).  The comparison
with chemical and chemodynamical evolution models is expected to provide insights into
the formation epoch of the different Galactic components and the
relation between them.

Chemical evolution models of the Milky Way usually divide the $"$Galaxy$"$ into
annuli with no radial transfer of material between them \citep[e.g.][]{FG03,
  Chiap03}.  However, recent works (e.g. Ro{\v s}kar et al.\ 2008b;
Sch\"onrich \& Binney 2008\nocite{SB08}; Haywood 2008\nocite{Hay08})
has pointed out the importance that migration mechanisms might have
in the studies of chemical evolution, in particular, upon the
observational constrains used to calibrated the models. For example, one of
the most difficult observational results to reproduce with current $"$semi-numerical$"$
chemical evolution models is the lack of an apparent correlation between
the age and the metallicity of the stars in the solar neighbourhood, as
well as the large scatter in metallicity at a given age.  On the other hand,
full chemodynamical models indicate that radial migration might be an
explanation to this lack of correlation.  \citet{Rok08b} have
shown that radial migration can largely affect the age and metallicity
gradients, as well as the dearth of metal-poor
stars in the solar neighbourhood (the so-called ``G-dwarf problem'').  
It has even been suggested that radial migration might be
responsible for the formation of the thick disc (e.g. Sch\"onrich \&
Binney 2008, Haywood 2008).

These studies provide the motivation to re-examine the stellar
population distribution within our simulated disc, where 
infall of material and the accretion of satellites are taken into account
naturally within our cosmological framework.

\subsubsection{The age-metallicity relation}
Chemical evolution models predict an increase of the metal content in
the ISM as stellar generations die and pollute their surroundings
with the byproducts of nucleosynthesis.
One might expect therefore that the youngest (oldest) stars would
also be the most metal-rich (metal-poor). 
However, there is little evidence of such an
age-metallicity relation in the solar neighbourhood \citep[][]{Fel01,
Nod04}. Furthermore, the scatter in the relation is very large --
i.e., at a given age, there is a large spread in the metallicity
distribution. Classically, this was interpreted as a result of
inefficient mixing of stellar ejecta.  However, several authors
\citep[e.g.,][]{Wie96,SB02} have suggested the possibility that radial
migration may have enabled old stars, formed at small galactocentric
radii, to appear in the solar neighborhood, flattening any correlation
between age and metallicity.  \citet{Rok08b} and Sch\"onrich \& Binney
(2008) have shown that a flatter age-metallicity relation and a large
spread in metallicities at a given age can be obtained with dynamical
models where radial mixing is taken into account. Haywood (2006,
2008)\nocite{Hay06} observed that stars in the metal-poor and metal-rich 
end of the thin disc have orbital parameters which are offset
from the main population, which he interpreted as being the
consequence of radial migration.

\begin{figure}
\resizebox{0.4\textwidth}{!}{\includegraphics[angle=-90]{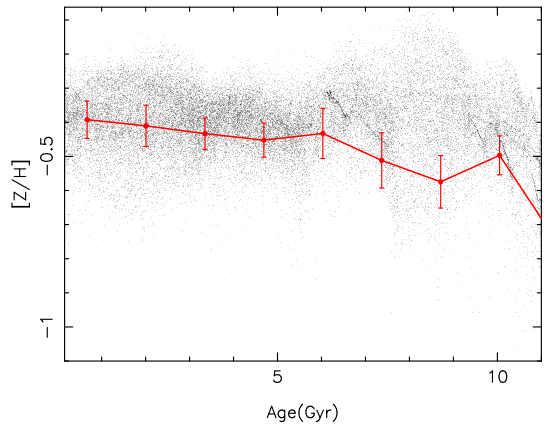}}
\caption{Age-metallicity relation for stellar particles with radius 
between 7 and 9~kpc and scale-height $|z| <3$ kpc, chosen 
to represent a region roughly corresponding to the ``solar neighbourhood''.
The red points indicate the values derived for the stars
born in this region.\label{agemet}}
\end{figure}

Fig.~\ref{agemet} shows the age-metallicity relation in the $"$solar
neighborhood$"$ (7 $<$ r $<$9 kpc and $|$z$|<$ 3 kpc) of our simulated disc,
 compared with the
age-metallicity relation that we would have measured in the absence of
migration.  As can be seen, our fully-cosmological study is in
agreement  with the aforementioned idealised models. 
Radial migration produces a considerable flattening, and an
increase of the scatter, in the age-metallicity relation. In the absence of
radial migration, the stars in the solar neighborhood of our disc
would show a relation between these two parameters.

\subsubsection{Age  gradient}

We showed, in Sec.~\ref{ageprofile} that the radial age profile has a
minimum at the break radius, similar to that found by \citet{Rok08b}.
In their work, the
radial extent of the star forming disc was limited by the maximum
angular momentum of material that was able to cool at each
timestep. Therefore, by construction, the disc grows inside-out and
the youngest stars are situated at the break radius. The $"$up-bend$"$ in the  age
profile at r$>$R$_{\rm br}$ was produced, exclusively, by stars that
had migrated from the internal parts of the disc. Because older stars
have more time to travel larger distances, the trend of the age with
radius reversed after the break radius.

\begin{figure}
\resizebox{0.4\textwidth}{!}{\includegraphics[angle=-90]{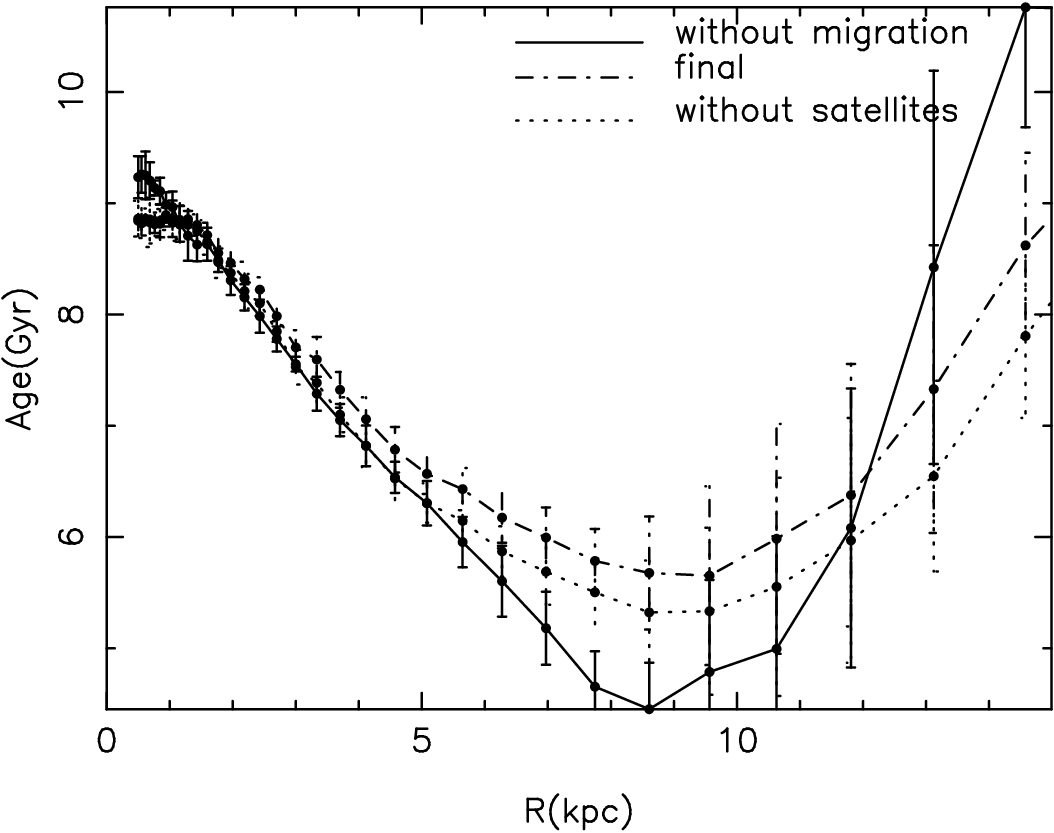}}
\resizebox{0.4\textwidth}{!}{\includegraphics[angle=-90]{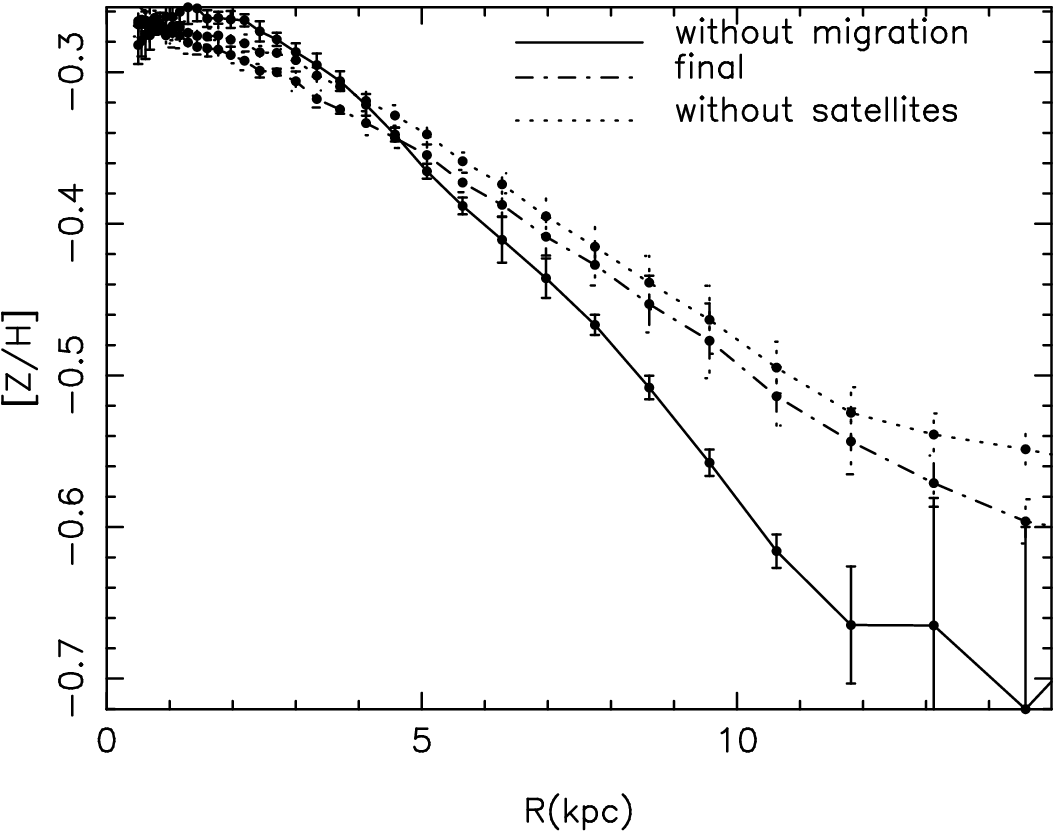}} 
\caption{Mass-weighted azimuthally-averaged stellar age and metallicity gradients.
Solid lines: Theoretical gradient for the hypothetical case were stars do not migrate from their birth place.
Dashed-dotted line: Profile measured in the final timestep of our simulation.
Dotted line: Profile measured in the final timestep of the simulation
after eliminating those stars which formed outside 
the disc (those with initial galactocentric radii in excess of
 25~kpc).
\label{age.metal.grad2}}
\end{figure}

\begin{figure}
\resizebox{0.4\textwidth}{!}{\includegraphics[angle=-90]{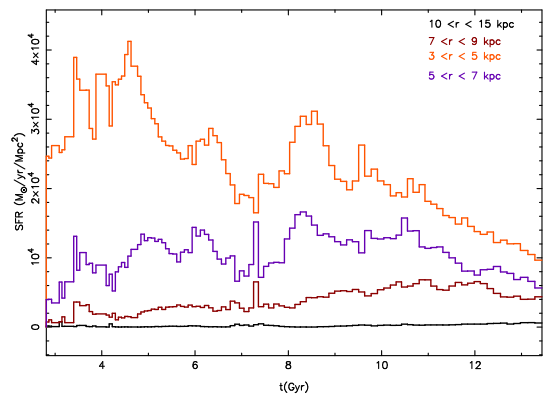}}
\caption{Comparison of the star formation rate density with time in 4 different regions of the 
galaxy disc -- black: between 10 and 15~kpc; dark red: between 7 and 9 kpc; purple: between 5 and 7 kpc
and orange between 3 and 5 kpc.\label{sfr}}
\end{figure}

\begin{figure}
\resizebox{0.5\textwidth}{!}{\includegraphics[angle=-90]{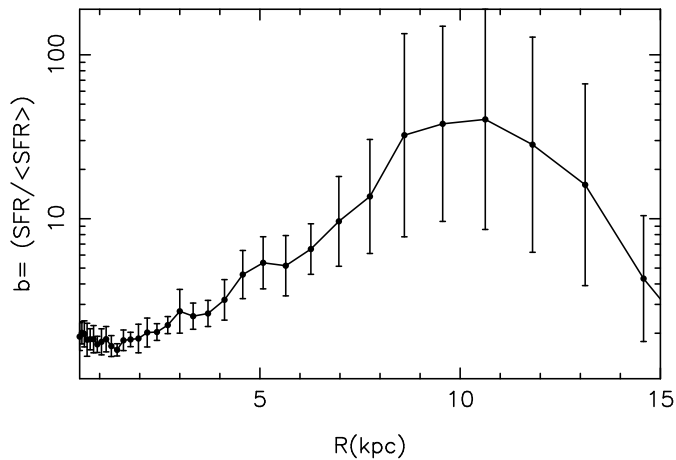}}
\caption{Star formation over the last 1~Gyr divided by the total mass of stars 
formed before this epoch (the so-called ``birth rate'') as a 
function of galactocentric radius.\label{birth}}
\end{figure}

To explore if this is also true in our simulation, we plot, in
Fig.~\ref{age.metal.grad2}, the mass-weighted age profile that the
stellar disc would have if the stars did not migrate from their birth
place, compared with that derived from the final at $z=0$.  As can be seen, even in the
absence of radial migration, a 'U-shape' profile in the age
gradient is still visible.  Stellar migration changes
the shape of this profile, but the increasing mass-weighted age with increasing 
radius in the outer parts is {\it not} due to radial migration.
 Figure~\ref{sfr} shows the evolution with time of the total
star formation rate density for stars formed at
different  galactocentric radii, including regions inside and beyond the 
break radius.  For radii 3$<$$r$$<$5~kpc, the SFR density 
decreases with time,
while the opposite happens for 5$<$$r$$<$7~kpc. The consequence 
of this is that the the fraction 
of young-to-old stars increases with radius until $r$=$R_{\rm br}$, 
consistent with expectations, as the
early accreted mass has low specific angular momentum.
These trends are responsible for the decreasing mean age with radius and are 
in agreement with the inside-out scenario for the formation of disc galaxies
(Ryder \& Dopita 1994). 
However, the trend dissapears beyond the break radius, 
where the star formation rate
is low and essentially constant throughout the evolution of the galaxy.
The difference in the star formation rate between the outer (10$<r<15$~kpc)
and inner (7$<$$r$$<$9~kpc) parts is higher at later times, and that is the main
reason for the upturn in the age gradient.
Another way to appreciate this is by studying the
radial profile of the birth parameter, defined as the current
versus averaged star formation rate $b$=(SFR/$<$SFR$>$)\citep{Kenn94}.
Instead of plotting the current star formation rate, we  averaged
the star formation over the last 1 Gyr.  This is mainly due to the relative few number of 
particles with ages below this value outside the break radius.  We
 plot, in Fig.~\ref{birth}, this parameter as a function of galactocentric
radius.  As can be seen, $b$ increases almost linearly
with radius, as expected for an inside-out formation scenario, until
the break radius, where it reaches a plateau and then decreases. 
 The increase
in the error bars at the break radius reflects the asymmetries in the age
distribution of the stars beyond this radius (recall Fig.~\ref{disk.sfr}).

We argue that the U-shape age profile  is the direct consequence
of the existence of a break in the star formation density. If the star
formation outside the break had not decreased suddenly, the age
 gradient would decrease, or remain constant, until the edge of the
optical disc.  This is supported by the result of \citet{Bak08} who
only found the tell-tale U-shaped colour profiles for galaxies
possessing a Type~II profile.
The galaxies with an essentially pure exponential profile within their sample
showed a plateau (and not an up-bend) in the colours at large radii.
This up-bending age profile does not mean that the disc did not form
inside-out. In fact, the ``overall'' formation of the disc remains inside-out. 
However, in our disc, the decrease of star formation in the external 
parts -- due to a decrease in the volume density of the gas -- 
results in redder colors beyond the break radius.

In Fig.~\ref{age.metal.grad2} we also compare the age profile of the
galaxy with the one it would have if {\it all} the stars formed in the
disc -- i.e., if satellites were not accreted. 
It is apparent that for this case, the accretion of satellites has little
effect on the stellar population gradients.

\subsubsection{Metallicity gradient}

The metallicity gradient in the disc and its evolution with time provides
constraints to our understanding of the formation and evolution
of galaxies.  The presence of a metallicity gradient in the Milky
Way is widely accepted, although its exact slope and shape
remain contentious \citep{Chiap01, And02}.  A related issue is
the evolution of this gradient with time; this has been approached
from both the theoretical (G\"otz \& K\"oppen~1992\nocite{GK92}; K\"oppen 1994\nocite{Koe94}; Molla et
al. 1997\nocite{Moll97}; Henry \& Worthey 1999\nocite{HW99}; Chiappini
et al. 2001\nocite{Chiap01}) and  observational perspectives (Friel et al.\
2002\nocite{Frie02}; Maciel 2001\nocite{Mac01}; Maciel, Costa \&
Uchida 2003\nocite{Mac03}; Stanghellini et al. 2006\nocite{Stan06})
but it is still not clear whether the metallicity gradient in our Galaxy
flattens or steepens with time. Measurements using HII regions,
B-stars, and planetary nebulae, find gradients ranging from $\sim$$-$0.04
to $\sim$$-$0.07 dex~kpc$^{-1}$ \citep[][although see Stanghellini et
al. 2006\nocite{Stan06}]{Dehar00, Affler97, Daf04, Gumm98, Cos04,
Mac99}.  Even more recently, \citet{Carr07}, using five old open clusters,
suggest an even shallower gradient of $-$0.018~dex~kpc$^{-1}$. There is 
also controversy in the literature about
the shape of the gradient, including the presence or not of a 
discontinuity at around $10-12$ kpc.
Part of the uncertainty in this field can no doubt be traced to the 
necessary use of disparate tracers of the gradients, each which
probe a different age (or range of ages).

\begin{figure}
\resizebox{0.4\textwidth}{!}{\includegraphics[angle=-90]{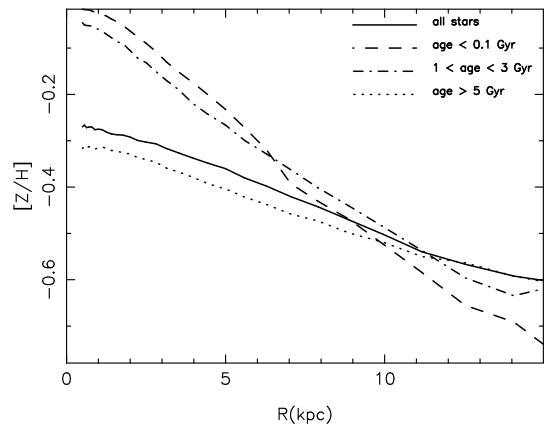}}
\caption{Azimuthally-averaged metallicity profile at redshift $z$=0 
for stars in different age bins, as noted in the inset.\label{meta.grad.age}}
\end{figure}

In our simulations, we find that the metallicity gradient
is flatter for old stars than it is for young stars, as shown in 
Figure~\ref{meta.grad.age}.  It can also be seen that radial 
migration is partially responsible for this 
flattening (see Fig.~\ref{age.metal.grad2}), in agreement with \citet{Rok08b}. 
We also predict that the optimal galactocentric radius in which to seek
differences between the gradients in the old versus young populations
is 3$<$$r$$<$5~kpc.
At can be seen in Figure~\ref{age.metal.grad2} (lower panel), 
the mean metallicity  gradient in our simulation shows a flattening at $r$$>$12~kpc.

\section{Discussion and Conclusions}
We have studied the origin of the disc truncations and the stellar population 
properties of a disc galaxy formed within a cosmological framework. 
To do this, 
we have analysed a multi-resolved cosmological simulation of a Milky Way-mass
halo, including 
dark matter and gas dynamics,  metallicity-dependent cooling, 
UV heating, star formation, and supernovae feedback,
using the AMR code {\tt RAMSES}.
Our simulated disc shows a break in the surface brightness
profile at $\sim$3 exponential scale-lengths, similar to those observed in a large
fraction of disc galaxies.  The position of the break does not
coincide with the radius at which the gas density reaches the canonical
threshold for star formation.  In our simulation, this break in the
stellar light is due to a decrease in the star formation
density per unit area (averaged over the past $\sim$1~Gyr).  This decrease in the star formation density originates from a
decrease in the volume density of gas at the break radius, which
itself coincides with the radius at which the gas disc begins to
warp (R$_{\rm warp}$).  A relationship between truncations and warps has
been pointed out
by several studies (see van der Kruit 2008\nocite{vdK08} for a review as well as van der
Kruit 2001\nocite{vdK01}).
van der Kruit (2007)\nocite{vdK07} 
found that the distribution of R$_{\rm warp}$/R$_{\rm br}$
in a sample of SDSS galaxies was 
statistically consistent with all warps starting at
$\sim$1.1~R$_{\rm br}$.  Our simulation is entirely consistent with
these observations.  
At a first sight, therefore, it seems that the presence
of a warp is a condition for the presence of a break in the stellar
light distribution. However, we re-iterate that this need to be confirmed with a
larger sample of simulated and observed discs.

We analysed the redistribution of material and angular momentum in the
disc, finding that it affects considerably the final shape of the mass
density profile, especially in the outskirts.  In fact, 57\% of the
stars with r$> R_{\rm br}$ formed at lower radii. 
The reason why the surface density profile does not show a truncation
similar to the one observed in the stellar light is traced to
this migration of
stars towards the outer disc.  We suggest that truncated 
and anti-truncated profiles can be produced by a different combination
of two processes: (1) a change in the slope of star formation profile
with radius due to a change of the slope in the gas density profile
(which causes the truncation in the light), and (2) radial migration of
stars formed in the internal parts towards radii in excess of $R_{\rm br}$.
 We speculate that up-bending profiles can be  produced in our simulation when the
distribution of the gas changes smoothly with radius. In that case,
process (1) is suppressed, the star formation rate per unit area in the
disc changes smoothly with radius, and the only deviations from a pure
exponential profile are due to stars migrating from disc interior,
producing an increase in both the surface light and mass density in
the external parts. This would make the external parts somewhat redder,
in agreement with observations (Bakos et~al. 2008).  Pure exponential
profiles can be similar to anti-truncated profiles, but with less
migration.

We have also studied the stellar populations in the disc and the
influence of the migration of stars and the accretion of satellites in
the age and metallicity profiles. We found, in agreement with recent
observations (Bakos et~al.\ 2008) that the disc shows a U-shaped age
profile, reaching the minimum age at the position of the break. It has
been proposed in recent works that this profile is produced by the
combination of two process: (1) the inside-out growth of the disc
(resulting in the age decreasing with radius until $R_{\rm br}$), and (2)
the migration of old stars from the interior to the exterior parts of
the disc beyond $R_{\rm br}$. In this picture, the increase of age
with radius beyond the break results from older stars having more
time to travel from the inner disc and, therefore,  reach a greater
galactocentric radius. These models predict a flat metallicity gradient for the old
stellar populations, as these stars are mixed very efficiently.  In our
simulation, the U-shaped profile
is due to a different rate of star formation between the
internal and the external regions (inside and outside the break
radius). Migration of the stars
also has an impact on  this profile,
but the U-shape age profile with the minimum  at the break radius
appears even when migration is  suppressed in the disc. We find
that the U-shaped age profile is due to the same mechanism as
the one producing the break in the surface brightness distribution -  ie,
a diminution of the star formation rate in the external region, 
due to the warping of the gaseous disc
and, therefore we predict that this type of color profile is only
present when there is a break in the light distribution.  This is in
agreement with the recent work of Bakos et al. (2008), at low
redshifts, and Azzolini et al. (2008) at redshifts out to $z\sim$1.1.

The distribution of ages in the outer disc of our simulation is very
broad with stars nearly as old as the age of the Universe itself.  The
presence of old stars ($>$10 Gyr) in the outer disc of some nearby
galaxies, such as M31 (Ferguson \& Jonhson 2001), has been used as an
argument against $\Lambda$CDM models, arguing that, if feedback is
included in order to produce large discs, the formation of the disc is
delayed and the resulting stellar populations should be of young and intermediate
ages. We show here that in a $\Lambda$CDM framework we were able to
produce disc with both a realistic scale-length and  a considerable
number of old stars in its external parts.  The metallicity
distribution of stars in the outskirts of the disc peaks at
[Fe/H]$\sim -0.5$ and shows also a significant spread, also
consistent with observations.

Finally, we studied the age-metallicity relationship in a representative
$"$solar neighbourhood$"$ of our simulation. 
There have been claims suggesting that radial mixing of stars
might be responsible for the absence of an obvious 
correlation between these two
parameters in the solar neighbourhood, and for the large scatter in metallicity at a given age.
 We agree with these claims, showing that a
flat relationship between age and metallicity, and significant scatter in metallicity at a given age, 
are a natural outcome within our
cosmological simulation.

In a forthcoming paper we will study the origin of the warp in the 
gaseous distribution, its lifetime, asymmetry, and intensity.  This
will be supplemented with a suite of 20-30 additional disc simulations,
spanning a range of mass, environment, and assembly history.  This
will allow us to provide significantly more robust predictions for
the relationship between break radii and other empirical characteristics
of galaxies.

\section*{Acknowledgments}

We wish to thank Romain Teyssier, Ignacio Trujillo, 
Armando Gil de Paz, and Isabel P\'erez, for their constant and extremely 
enlightening advice. We thank Ruyman Azzolini for providing his data
in tabular form and the referee of this paper, Dr Peter Erwin for useful 
comments that have improved the presentation of the manuscript.
We also thank Victor Debattista and Rok Ro{\v s}kar for useful comments that
improved significantly the manuscript.
PSB acknowledges the support
of a Marie Curie Intra-European Fellowship within the 6th
European Community Framework Programme.
BKG acknowledges the support of the UK's Science \& Technology
Facilities Council (STFC Grant ST/F002432/1) and the
Commonwealth Cosmology Initiative. 
All modeling and analysis was carried out on the University of Central
Lancashire's High Performance Computing Facility and the UK's
National Cosmology Supercomputer (COSMOS).
\bibliographystyle{mn2e}
\bibliography{references}

\appendix

\section{Dependence of the break radius upon inclination}
 
Recently, van der Kruit (2008) studied the correlations between break 
radius and other galaxy properties, finding fundamental differences 
between the trends described by face-on and edge-on galaxies. In 
particular, he did not find a correlation between 
the position of the break and the rotational velocity in the 
face-on sample, while  the correlation was evident in the edge-on galaxies.
In principle, breaks are more easily detected in edge-on galaxies, due
to their favored integrate line-of-sight vantage point.
Face-on breaks are more difficult to detect and if the truncation 
radius changes with azimuth, the azimuthally-averaged profiles 
would ``smear out'' the intensity of the break.
Furthermore, intrinsic deviations from circular symmetry, such 
as spiral arms (especially for 
younger populations) should complicate matters for less-inclined systems.
On the other hand, edge-on studies are more susceptible to
line-of-sight effects caused by integrating through the disc.
The differences between the parameters measured in face-on and 
edge-on samples can be 
obtained statistically, but obviously, they are difficult to 
quantify. In this appendix,
we compare the break radius and the intensity of the break in different 
bands by viewing our simulation various face-on and edge-on configurations,
in order to quantify possible differences due to geometry.
Differences due to the varying dust contribution with inclination
cannot be treated here, but we refer
the reader to Pohlen et al. (2008) for a thorough analysis of this issue.

\begin{figure}
\resizebox{0.4\textwidth}{!}{\includegraphics[angle=-90]{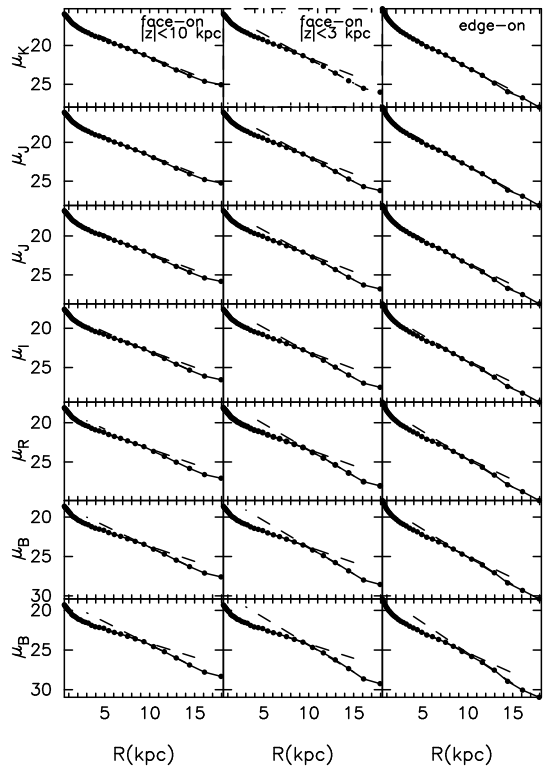}}
\caption{
Surface brightness profiles in different bands calculated
in face-on (azimuthally averaging)
and edge-on configurations. For the face-on projection we have 
integrated between different distances from the disc mid-plane, 
as indicated in the inset.\label{inclination}}
\end{figure}

All the profiles previously showed in this work have been obtained by
azimuthally-averaging the luminosities
of the stars within 3~kpc of the disc mid-plane. Observers, of course, 
integrate their 
luminosities along the entire 
line-of-sight. In order to mimic this we compare the profiles integrating
the light of all stars within a distance of 10~kpc from the disc mid-plane.
Figure\ref{inclination} shows the surface brightness profiles in 
different bands obtained for face-on and 
edge-on orientations.
Table~\ref{tab:inclination} shows the measured values for the position of the break 
and the angle between the 
two exponential fits in each configuration. 

We do not find any fundamental difference between the position of the 
break or its intensity 
for the face-on and edge-on values; they agree, within the uncertainties.
The table also shows the measured inner and outer scale-lengths. 
The uncertainties in the measurement 
of the scale-length are $<$0.1\% and are, therefore, not listed.
We can see that contrary to what happens with the position of the 
break, we did find 
differences in the scale-lengths of the disc when measured at different 
inclinations. 
In particular, the scale-lengths are systematically lower in the 
edge-on projections, counter to that claimed by
van der Kruit (2008).
Due to the smaller scale-length obtained in edge-on systems, the 
$R_{\rm BR}/h$ measured 
for this projection is larger than for the face-on projection.
These differences have been found previously in empirical 
studies, but it is not clear if they 
are due to projection effects or to different techniques to mark the break (see 
Trujillo \& Pohlen 2006).

Finally, we do not find a sharper edge in edge-on galaxies than for 
face-on systems, for which we conclude
that this effect, reported in observations, must be the consequence of 
dust extinction.

\begin{table}
\begin{tabular}{lcccc}
\hline\hline
           & band      &face-on           & face-on        & edge-on         \\
           &           & $|$z$|<$10 kpc       & $|$z$|<$ 3 kpc     &                 \\  
\hline
R$_{\rm br}$&  B        &$9.3\pm 0.3$      &   $9.6\pm0.7$  &  $9.8\pm 1.0$   \\  
h$_{inn}$   &  B        & 2.83             & 2.75           &  1.95  \\
h$_{out}$   & B         & 1.74             & 1.46           &  1.30  \\ 
angle      &  B        & 10.2             &  14.4          & 10.4            \\
\hline
R$_{\rm br}$ &  V       & $9.2\pm 0.4$      &  $9.5\pm0.6$   &  $9.3\pm 1.3$   \\
h$_{inn}$   &   V       & 2.61              &  2.53           &  1.81 \\
h$_{out}$   &   V       & 1.83              &  1.47           &  1.41 \\
angle       &  V       &  7.6             &  12.5           &  6.3    \\
\hline
R$_{\rm br}$ &  R       & $8.9\pm0.7$      &  $9.8\pm0.4$   &  $9.2\pm1.7$    \\
angle       &  R       &  5.9             &  11.1          &   6.5           \\
h$_{inn}$   &   R       &  2.51            &  2.43          & 1.83          \\
h$_{out}$   &   R       &  1.90            &  1.52          &  1.42          \\
\hline     
R$_{\rm br}$ &  I       & $8.7\pm1.1$      & $9.6\pm0.4$    &$8.3\pm1.0$   \\
angle       &  I       &   4.6            &  9.3           &  4.2         \\
h$_{inn}$   &   I       &   2.42           & 2.34           & 1.80        \\
h$_{out}$   &   I       &   1.96           & 1.57           & 1.52        \\
\hline
R$_{\rm br}$&    J      & $8.2\pm1.7$      & $9.5\pm0.3$    &  $7.8\pm1.4$    \\
angle      &    J      & 3.4              &  8.2           &   3.4           \\
h$_{inn}$   &   J       &  2.31            &  2.24          &  1.76         \\
h$_{out}$   &   J       &  1.99            &  1.58          &  1.54          \\
\hline
R$_{\rm br}$&     H     & $8.1\pm2.9$      &$9.5\pm0.3$     &  $12.1\pm 6.0$ \\
h$_{inn}$   &     H     & 2.20             & 2.20           &  1.60\\
h$_{out}$   &     H     & 2.00             & 1.59           &  1.52\\
angle      &     H     & 2.1              &  7.7           &   1.3          \\
\hline
R$_{\rm br}$&    K     & $9.2\pm2.1$      & $9.4\pm0.3$    &    --       \\
h$_{inn}$   &    K     & 2.22            & 2.19            &  1.65       \\
h$_{out}$   &    K     & 1.95            & 1.58            &  1.55       \\
angle      &    K     & 2.8              &  7.7           &    0        \\
\hline
\end{tabular}
 \caption{Break radius and angle between the two exponential fits in different bands.\label{tab:inclination}}
\end{table}
\label{lastpage}

\end{document}